\documentclass{statsoc}

\usepackage[a4paper]{geometry}
\usepackage{graphicx}
\usepackage{amsmath}
\usepackage{natbib}

 \RequirePackage[colorlinks,citecolor=blue,urlcolor=blue]{hyperref}

\usepackage{amsmath, amsfonts, amssymb, multirow}

\newtheorem{lemma}{\textbf{Lemma}}
\newtheorem{theorem}{\textbf{Theorem}}
\newtheorem{definition}{\textbf{Definition}}

\newtheorem{prop}{\textit{Proposition}}

\newtheorem{algo}{\textit{Algorithm}}

\newcommand{\usmall}{u}
\newcommand{\vsmall}{v}
\newcommand{\Ut}{U_t}
\newcommand{\Vt}{V_t}

\newcommand{\U}{U}
\newcommand{\Uz}{U_0}

\newcommand{\Un}{U_n}
\newcommand{\Uton}{U_{0:n}}

\newcommand{\Ui}{U_{i}}
\newcommand{\Uim}{U_{i-1}}
\newcommand{\Uip}{U_{i+1}}
\newcommand{\Utoi}{U_{0:i}}
\newcommand{\Utoim}{U_{0:i-1}}
\newcommand{\V}{V}

\newcommand{\Vton}{V_{0:n}}
\newcommand{\Vz}{V_0}

\newcommand{\Vi}{V_{i}}
\newcommand{\Vim}{V_{i-1}}
\newcommand{\Vip}{V_{i+1}}
\newcommand{\Vtoi}{V_{0:i}}
\newcommand{\Vtoim}{V_{0:i-1}}
\newcommand{\pDelta}{p_\Delta}

\newcommand{\EDelta}{\mathbb{E}_\Delta}
\newcommand{\E}{\mathbb{E}}

\newcommand{\hthetamm}{\widehat{\theta}_{m-1}}
\newcommand{\hthetam}{\widehat{\theta}_{m}}
\newcommand{\hthetaz}{\widehat{\theta}_{0}}

\newcommand{\Weighttonk}{W^{(k)}_{0:n}}

\newcommand{\Weightz}{W_0}
\newcommand{\weightz}{w_0}
\newcommand{\Weighti}{W_i}
\newcommand{\Weight}{W}
\newcommand{\Weightn}{W_n}
\newcommand{\weighti}{w_i}
\newcommand{\Weightim}{W_{i-1}}

\newcommand{\Utonk}{U^{(k)}_{0:n}}
\newcommand{\Utoik}{U^{(k)}_{0:i}}

\newcommand{\Uik}{U^{(k)}_{i}}
\newcommand{\Uzk}{U^{(k)}_0}

\newcommand{\R}{\mathbb{R}}

\newcommand{\X}{\mathcal{X}}

\title[Inference for hypoelliptic diffusions]{Hypoelliptic diffusions:
filtering and inference from\\ complete and partial observations}
\author[Ditlevsen and Samson]{Susanne Ditlevsen}
\address{Department of Mathematical Sciences,
  University of Copenhagen.}
\email{susanne@math.ku.dk}
\author[Ditlevsen and Samson]{Adeline Samson}
\address{Univ. Grenoble Alpes, LJK, F-38000 Grenoble, France\\
              CNRS, LJK, F-38000 Grenoble, France.}
\email{adeline.leclercq-samson@imag.fr}

\begin{document} 
\maketitle

\begin{abstract}
The statistical problem of parameter estimation in partially observed
hypoelliptic diffusion processes is naturally occurring in many
applications. However, due to the noise structure, where the noise
components of the different coordinates of the multi-dimensional
process operate on different time scales, standard inference tools are
ill conditioned. In this paper, we propose to use a higher order
scheme to 
approximate the likelihood, such that the different time
scales are appropriately accounted for. We show consistency and
asymptotic normality with non-typical convergence rates. When only
partial observations are available, we embed the approximation into a
filtering algorithm for the unobserved coordinates, and use this as a
building block in a Stochastic Approximation Expectation Maximization
algorithm. We illustrate on simulated data from three models; the
Harmonic Oscillator, the FitzHugh-Nagumo model used to model the
membrane potential evolution in neuroscience, and the Synaptic
Inhibition and Excitation model used for determination of neuronal
synaptic input. 
\end{abstract}

\section{Introduction}

Hypoelliptic diffusion processes appear naturally in a variety of
applications, but most parameter estimation procedures are ill
conditioned, especially when only partial observations are available. 
Hypoellipticity means that the diffusion matrix of the
stochastic
differential equation (SDE) defining the multidimensional diffusion process is not of full
rank, but its solutions admit a smooth density. In this paper we
consider parametric estimation for hypoelliptic 
diffusions defined as solutions to an SDE of the following form:  
\begin{equation}\label{GeneralModel1}\left\{
\begin{array}{lcl}
dV_t&=&a(V_t,U_t) dt\\
dU_t&=&A(V_t,U_t)dt+\Gamma(V_t,U_t)dB_t 
\end{array}
\right.
\end{equation} 
where $ V_t\in\mathbb{R}$ and $U_t\in \mathbb{R}^p$, from discrete
observations of the full system $(V_t,U_t^T)^T$, or from discrete
observations of $V_t$ only (partial observations), the latter being the
most realistic in applications. Here, $^T$ denotes
transposition. 
{The components of $U_t$ are {\em rough}, since the
noise acts directly on $U_t$, whereas $V_t$ is only 
indirectly affected by the noise. The noise is propagated through
$a(\cdot)$, which has to
depend on $U_t$ for the model to be hypoelliptic, and thus, $V_t$ is
the {\em smooth} component. }

A prominent example is the
large class of stochastic damping Hamiltonian systems, also called
Langevin equations, describing the motion of a particle subject to 
potential, dissipative and random forces 
\citep{Wu2001,Cattiaux2014,Cattiaux2014a, Cattiaux2015,Comte2017}. In this case $a(\cdot) =
U_t$ and $A(\cdot) = -c(V_t,U_t)U_t - \nabla P(V_t)$, for some function
$c(\cdot)$ and where $P(\cdot)$ is
the potential. They typically arise from a
second order differential equation, which develops into a higher
dimensional system with some coordinates representing positions, and
some coordinates representing velocities. The noise is degenerate
because it acts directly on the coordinates of 
the momentum only, and not on the positions. These models have many
applications, such as molecular dynamics
\citep[eqs. (6.30)-(6.31)]{LeimkuhlerMatthewsBook2015}, stochastic
volatility models, paleoclimate research \citep{Ditlevsen2002}, neural mass models 
\citep{Hinterleitner2017}, random mechanics or classical 
physics. Specific examples are the harmonic oscillator (HO), where
$A(\cdot) = -DV_t -\gamma U_t$, which will be our first example, the
van der Pol oscillator where $A(\cdot) = \mu (1-V_t^2)U_t -V_t$ and
the Duffing oscillator where $A(\cdot) = -\delta U_t 
-\beta V_t - \alpha V_t^3+\gamma \cos \omega t$. In this 
setting, parametric estimation has been considered before, taking
advantage of the special structure of $a(V_t,U_t)=U_t$. 
\citet{Samson2012} propose contrast estimators based on the fully
observed system, by approximating the unobserved coordinate $U_t$ by the
{increments} of the observed coordinate $V_t$. \citet{Pokern2009} propose a Gibbs
algorithm in a Bayesian framework, still relying on the simple form of
$a$. The
particular case of integrated diffusions, where the dynamics of $U_t$
do not depend on $V_t$, has been investigated by 
\citet{genon-catalot2000,Ditlevsen2004,Gloter2006}. 

However, many applications need to allow for a more flexible
formulation of {the} function $a(\cdot)$. For example, it can be convenient to model
parts of a large deterministic system exhibiting multiple time scales
by a low dimensional stochastic model, leading to a hypoelliptic structure
on the reduced model \citep{PavliotisStuartBook}. An important field of application
is neuronal models of membrane potential evolution, where the noise
only acts on the input, or on the ion channel dynamics, leading to
hypoelliptic SDEs. Examples are the FitzHugh-Nagumo (FHN) model
\citep{DeVille2005,LeonSamson2017}, which is our second example, the Hodgkin-Huxley model 
\citep{GoldwynSheaBrown2011,TuckwellDitlevsen2016}, or conductance based models with
stochastic channel dynamics \citep{DitlevsenGreenwood2013}. Also
neural field models are often hypoelliptic
\citep{CoombesByrne2017,DitlevsenLocherbach2017}. It is therefore
important to develop reliable estimation methods for this class of
models. A particular sub-class are hypoelliptic homogeneous
Gaussian diffusions, where the drift is linear and the diffusion is
constant, which were considered by \citet{A1985}, and where the
transition density is explicitly known. A simple example is
the HO mentioned above. 

Ergodicity of these models has been studied, based on the
hypoellipticity of the system \citep{Mattingly2002}. But even if the
model is ergodic, the degenerate noise structure complicates the statistical
analysis and many standard tools break down. The main difficulty with
hypoelliptic models compared to the elliptic 
case is the transition density for time $\Delta$, which converges {pointwise} towards a point
measure when $\Delta \rightarrow 0$ at a faster rate {(with a 1-norm)}, $1/\Delta^2$
\citep{Cattiaux2014a,Comte2017}, compared to the elliptic case of $1/\Delta$. 
 In general, the transition density is
unknown, and the estimation fails if { the likelihood
  is approximated} by the
Euler-Maruyama scheme, since the scheme can fail to be ergodic for any
choice of time step, even
if the underlying SDE is \citep{Mattingly2002}. Intuitively, the problem arises because the
diffusion matrix is not of full rank, and lower order schemes will
have a degenerate variance matrix, even if the underlying model does
not, {due} to the hypoellipticity. As a simple example consider an
integrated Brownian motion $dV_t = U_t dt; dU_t = \sigma dB_t$. The
exact transition density is normal,
\begin{eqnarray*}
\left ( \begin{array}{c}  V_{\Delta} \\ 
    U_{\Delta} \\ \end{array} \right ) \sim N \left (\left ( \begin{array}{c}
                                                        V_0 + U_0 {\Delta} \\ 
    U_0 \\ \end{array} \right ),\sigma^2\left ( \begin{array}{cc}
                                                  \frac{{\Delta}^3}{3} & \frac{{\Delta}^2}{2} \\ 
    \frac{{\Delta}^2}{2} & \Delta \\ \end{array} \right ) \right ),
\end{eqnarray*}
{with} a non-degenerate covariance matrix. However, if the
transition density is approximated by the Euler-Maruyama scheme, the
approximated transition density becomes
\begin{eqnarray*}
\left ( \begin{array}{c}  V_{\Delta} \\ 
    U_{\Delta} \\ \end{array} \right ) \sim N \left (\left ( \begin{array}{c}
                                                        V_0 + U_0 {\Delta} \\ 
    U_0 \\ \end{array} \right ),\sigma^2\left ( \begin{array}{cc}
                                                  0 & 0 \\ 
    0 & \Delta \\ \end{array} \right ) \right ),
\end{eqnarray*}
which has a non-invertible covariance matrix, so the likelihood
function is not well defined.
 
\citet{Pokern2009} suggest to
circumvent this problem by adding the first non-zero noise terms
arising in the smooth components of the It{\^o}-Taylor expansion of
the process { corresponding to a weak order 1.5
  scheme. The covariance matrix then becomes the exact covariance
  matrix for the integrated Brownian motion above, which is also used
  as an approximation of the covariance matrix in more complicated
  models. Then they combine it with an Euler scheme for the inference
  of the drift in a Gibbs loop. They also show that using the weak
  order 1.5 scheme for inference of the drift parameters leads to a
  biased drift estimate}. 
 Instead we suggest to
approximate the unknown transition density with a higher order scheme,
namely the strong order 1.5  Taylor scheme \citep{KloedenPlatenBook},
which leads to the same approximation of the variance up to leading
order as in \citet{Pokern2009}, but also approximates the mean
{up to sufficiently high order}. We propose a contrast based on this scheme, and prove
consistency under the standard asymptotics of $\Delta \rightarrow 0$
and $n \Delta \rightarrow \infty$. 
The proof 
relies on the higher order approximation of the mean, and thus,
provides an explanation of why the consistency failed for the 
{weak order 1.5} estimator {of the drift parameters}  proposed by
\citet{Pokern2009}. To our surprise, we also
obtain asymptotic normality, but with faster convergence rates of
parameters of the smooth components than the usual rates of the rough
components. 

When only partial observations are available, i.e., only some
coordinates are observed, the statistical difficulties
increase. {The problem belongs to the class of
  state-space or hidden Markov models \cite[see for
  example][]{cappe05, Kantas2015}, but in a degenerate way. The
  degeneracy arises for two reasons.} One 
problem is that the system is coupled, such that the
unobserved coordinates are not autonomous, and the hidden Markov model
 is {the vector $(V_t, U_t)$, such that the
   distribution of the observations conditionally on the Markov process is
   being reduced to a (non-smooth and degenerate) Dirac
   density. Second, the variance of the discrete hidden Markov process
   is itself degenerate if the discretization is applied with a naive
   scheme.} We therefore embed the approximation into a filtering 
algorithm for the unobserved path and a Stochastic Approximation
Expectation Maximization (SAEM) algorithm, as suggested in \citet{Ditlevsen2014} for the
elliptic case.  This framework furthermore extends the class we can
handle considerably by allowing for general drift functions also for
the smooth components, as well as for state dependent diffusion matrices.

The running examples throughout the paper are the HO model, where we
compare with the estimators 
proposed in \citet{Pokern2009} and \citet{Samson2012}, the FHN model, where we
allow for a general $a(\cdot)$ in the drift of the smooth component,
and the Synaptic Inhibition and Excitation (SIE) model, where $p>1$
and the diffusion matrix is state dependent. In Section
\ref{sec:models} we introduce the general model, the likelihood and
notation, we discuss conditions for hypoellipticity, give formulas for
moments and introduce the three example models. In
Section \ref{sec:discretization} we give the discretization scheme and
present some theoretical results of the scheme needed to show consistency of the
estimators. In Section \ref{sec:bothobserved} we present contrast estimators for
the completely observed case, which will serve as a basis for the
partially observed case, where the unobserved components have to be
imputed before employing the contrast estimator. In Section
\ref{sec:unobserved} we introduce the particle filter to impute the
hidden path and the SAEM algorithm to estimate by alternating between
imputation and estimation from the fully observed system, and we give
indications of how to choose the 
initial parameter values for the algorithm. In Section
\ref{sec:simulation} we conduct a simulation study on the three
example models, and we compare with other
estimators. {Proofs are gathered in the Supplementary material}.

\section{Models}
\label{sec:models}

In this paper we consider parametric estimation for hypoelliptic
diffusions defined as solutions to an It\^{o} SDE of the following form: 
\begin{equation}\label{GeneralModel}\left\{
\begin{array}{lcl}
dV_t&=&a(V_t,U_t; \psi) dt\\
dU_t&=&A(V_t,U_t;\varphi)dt+\Gamma(V_t,U_t; \sigma)dB_t 
\end{array},
\right.
\end{equation} 
where $ V_t\in \X_V \subset\mathbb{R}$, $U_t\in \X_U \subset\mathbb{R}^p$ with $p\geq 1$ and
$B_t$ is a {$p$}-dimensional Brownian motion. Denote the full state
space by $(V_t,U_t^T)^T \in \X  \subset\mathbb{R}^{p+1} $. 
{The} functions  $a: \X
\mapsto \mathbb{R}$ and $A : \X
\mapsto \mathbb{R}^p$   are drift   functions depending on an unknown
parameter vector  $\beta=(\psi, \varphi)$. Denote the full drift vector by $b=(a,A^T)^T$. Furthermore, $\Gamma :\X 
\mapsto \mathbb{R}^{p\times {p}}$ is a partial diffusion coefficient
matrix depending on an 
unknown parameter vector $\sigma $, the full diffusion matrix being 
\begin{equation}\label{DifMatrix}
{C} (v,u;\sigma)=\left [ \begin{array}{c}  \mathbf{0}_{{ p}} \\ 
    \Gamma (v,u;\sigma) \\ \end{array} \right ],
\end{equation}
where $\mathbf{0}_{{ p}}$ is the {$p$}-dimensional row vector of
zeros. Equation \eqref{GeneralModel} is assumed to have a weak 
solution, and the coefficient functions $a, A$ and $\Gamma$ are
assumed to be smooth enough to ensure the uniqueness in law of the
solution, for every $\beta$ and $\sigma$. {Furthermore, the solution is
assumed to be ergodic.}  Most importantly, the
process is assumed to be hypoelliptic, meaning that it admits a smooth
density with respect to the Lebesgue measure, see Section \ref{sec:hypo}. 
{We assume diagonal noise, such that}
\begin{equation}\label{GeneralGamma}
\Gamma (v,u;\sigma)=\left [ \begin{array}{ccc} \sigma_1 (v,u;\sigma)& 0 & 0\\ 
0 & \ddots & 0 \\
0&0 & \sigma_p  (v,u;\sigma)\\ \end{array} \right ],
\end{equation}
where $\sigma_j (v,u;\sigma)>0$ for $(v,u^T)^T\in \X$ and $j=1, \ldots , p$.  
In the applications below $p=1$ or 2. 

\subsection{Likelihood and objectives}
In model (\ref{GeneralModel}), the parameters $\psi, \varphi$ and
$\sigma$ are unknown. The objective of this paper is to estimate these
from observations of the first coordinate $V_t$ at discrete times
$t_0, t_1, \ldots, t_n$, with equidistant time steps
$\Delta=t_{j+1}-t_j$. The ideal would be to maximize the likelihood
$p(\Vton; \beta, \sigma)$ of the data  $\Vton=(V_{0}, \ldots,
V_{n})$, where we write $V_j := V_{t_j}$ for $j=0,1,\ldots , n$.  However, the likelihood is intractable, not only because the
transition density of model \eqref{GeneralModel} is generally unknown,
but also because $\Vton$ is not Markovian, only $(V_t, U_t)$ is
Markovian. Even if there is no noise on the first coordinate, the
hypoellipticity condition implies that the transition 
 density of model \eqref{GeneralModel} exists. Denote the unknown transition
 density by $p(V_{t+\Delta},
 U_{t+\Delta}|V_t,U_t;\beta, \sigma)$, then the complete likelihood,
 assuming all coordinates {are}
 observed and using the Markov property of
 $(V_t, U_t)$, is given by
 \begin{equation}\label{eq:completeLikelihood}
 p(\Vton,\Uton; \beta, \sigma) = \prod_{i=0}^{n-1} p(\Vip, \Uip|\Vi, \Ui;\beta, \sigma).
 \end{equation}
The marginal likelihood of $\Vton$, when only the first coordinate is
observed, is a high-dimensional integral, 
  \begin{equation}\label{eq:Likelihood}
 p(\Vton; \beta, \sigma) = \int \prod_{i=0}^{n-1} p(\Vip, \Uip|\Vi, \Ui;\beta, \sigma)d\Uton,
 \end{equation}
which is difficult to handle. 

A standard approximation to the unknown transition density is given by
the Euler-Maruyama 
scheme, where the {true} transition density is {approximated by the Euler} normal {density} with mean and variance
given by the drift and diffusion coefficients multiplied by
$\Delta$. However, since the diffusion coefficient on the first
coordinate is zero, the normal distribution of the scheme is singular, and the
estimation breaks down. The same happens for the 
Milstein scheme{, which has strong order 1, compared to the
  Euler-Maruyama scheme, which has strong order $1/2$}. We suggest instead to approximate with a
higher order scheme with strong order 1.5, where, as we shall see, a
stochastic term of order $\Delta^{3/2}$ appears in the first coordinate, which
is a {smoothed} version of the stochasticity from the other
coordinates. This stochasticity is enough to ensure that the
estimation procedure 
works, as long as drift terms of the same order {in
  $\Delta$ are} maintained in the
approximation. Denote by 
\begin{equation}\label{eq:deltap}
 p_{\Delta}(\Vip, \Uip|\Vi, \Ui;\beta, \sigma)
\end{equation}
the approximated transition density from this scheme.

In Section \ref{sec:bothobserved}, we assume all 
coordinates $(V_t, U_t^T)^T$ {are} observed at discrete time points, and explain
how we can estimate the parameters in that case.  In Section
\ref{sec:unobserved} we assume only $\V_t$ observed, and 
suggest to impute the hidden 
coordinates $U_t$ and discuss how to maximize the likelihood $p_{\Delta}(\Vton;
\beta, \sigma)$.  Before detailing the estimation approaches, we give further details on hypoellipticity and some moment properties of the process. Section \ref{sec:discretization} is devoted to the discretization scheme of order 1.5. 

\subsection{Notation}

 Let $\bar \Gamma (v,u)= (\sigma_1 (v,u), \ldots ,
 \sigma_p (v,u))$ denote the vector of entries in the diagonal of matrix
 \eqref{GeneralGamma}. Let $\partial_{u}
 a (\Vi,\Ui)$ denote the row vector of partial derivatives evaluated
 at time $t_i$, $(\partial_{u_1}
 a (v,u), \ldots , \partial_{u_p}
 a (v,u))|_{(v,u)= (\Vi,\Ui)}$, and likewise for the Jacobian matrix
 of $A$ and {$\bar \Gamma$}. Let
 $\bigtriangledown^2_{\bar \Gamma} (\cdot )=\sum_{j=1}^p 
 \sigma_j^2(v,u)\frac{\partial^2}{\partial u_j^2}(\cdot)$ denote a weighted
 Laplace type operator. 
It is applied
 componentwise to vectors.  
Let $\partial_x f^i$ denote the $n$-dimensional
 row vector of 
partial derivatives of the $i$th component of a generic function $f: \X \to \R^{n}$ with respect to the
elements of $x$, or {write} $\partial_x f$ if $n=1$. We will sometimes
use notation $b$ for the drift, and sometimes $a, A$, depending on
what is most notational{ly} convenient. Note that $b_1=a$ and $b_{j+1}=A_j$ for $j=1,
\ldots , p$. We sometimes write $X_t=(V_t, U_t^T)^T$ for the process,
but use $V_t$ and $U_t$ when we need to distinguish between the smooth
and the rough parts of the process. Let $I_p$ denote the identity matrix of dimension $p$ 
 and $\mathbf{1}_p$ the $p$-column vector of ones.

\subsection{Hypoellipticity}
\label{sec:hypo}

An SDE is hypoelliptic if the {squared} diffusion matrix {$C
C^T$} is not of full rank,
but its solutions admit a smooth {transition} density with respect
to the Lebesgue measure. 
H{\"o}rmander's theorem asserts that this is the case if the SDE in
its Stratonovich form 
satisfies the weak H{\"o}rmander condition
{\citep{NualartBook}}. We write $\sigma^j: \R^{p{+1}} \to
\R^{p} $ for the ${p}$ column 
vectors of the diffusion matrix $\Gamma$, and $\tilde \sigma^j:
\R^{p+1} \to \R^{p+1} $ for the ${p}$ column 
vectors of the diffusion matrix \eqref{DifMatrix}, such that $\tilde
\sigma^j = (0, (\sigma^j)^T)^T$.

For smooth vector fields $f(x)$ and $ g( x) : \R^n   \to \R^n , $ the
$i$th component of the Lie bracket $ [f, g ] $ is defined by
$ [f, g ]^i = (\partial_x g^i) f - (\partial_x f^i) g \; , \; i = 1,
\ldots , n ${, where $(\partial_x g^i) f$ is the scalar
  product between the row vector $\partial_x g^i$ and the column
  vector $f$, and likewise for the second term}. 
Define the set ${\cal L}$ of vector fields by the initial members
$\tilde \sigma^j  \in {\cal L}, j=1, \ldots , {p}$ and recursively by 
\begin{equation}\label{eq:iteration}
L\in {\cal L} \;\Longrightarrow \; [ b,L] , [ \tilde\sigma^1  , L], \ldots,  [\tilde\sigma^{{p}} , L ]  \in {\cal L}  \;.   
\end{equation}
The weak H\"ormander condition is fulfilled if 
the vectors of ${\cal L}$ span $\R^{p+1}$ {for each $x \in \R^{p+1}$}. {The initial
members span {$\{(0,v)\in R^{p+1}  :  v \in
R^p \}$, a subspace of dimension $p$}, since $\Gamma (v,u)$
is given by \eqref{GeneralGamma}}. Therefore, we only need to check if
there exists some 
$L\in  {\cal L}$ which has the first element different from zero. The first iteration of
\eqref{eq:iteration} for system \eqref{GeneralModel} yields
\begin{align*}
\label{Liebrackets}
[b,\tilde\sigma^j]^1 &=  -\partial_{u} a(v,u) \sigma^j (v,u) \\
[\tilde\sigma^i,\tilde\sigma^j ]^1 &= 0
\end{align*}
for $i,j = 1, \ldots , p$. {If the first of these is 0,
all subsequent iterations will be 0.} This leads us to the following
sufficient and necessary condition for system 
\eqref{GeneralModel} to be hypoelliptic.
\begin{itemize}
\item[{\bf (C1)}] $\forall (v,u^T)^T \in \X , \, \partial_{u} a(v,u) \sigma^j (v,u) \neq 0$ for at least one $j=1, \ldots , p$.
\end{itemize}
This is a natural assumption; the noise on some of the components of $u$ should be propagated to the
first coordinate, which can only happen if $a(v,u)$ depends on at
least one component of $u$. Note that the system has to be in
its Stratonovich form, whereas we assume model \eqref{GeneralModel} in
its It\^{o} form. However, the condition still holds, since it
only involves the drift of the first component. If $\Gamma (v,u)$
in \eqref{GeneralGamma} does not depend on $(v,u^T)^T$, the It\^{o} and the
Stratonovich forms coincide. If it is state dependent, a conversion
from It\^{o} to Stratonovich form will change the drift functions of
the $U_t$ coordinates, but not of $V_t$.

\subsection{Moments}

The distribution of $X_t=(\Vt,\Ut^T)^T$ in eq. \eqref{GeneralModel} is
in general unknown, but moments can be
approximated when $X_t$ is ergodic. For sufficiently smooth and
integrable functions $f :\X 
\mapsto \mathbb{R}$ {(with respect to the invariant
  measure of $X$, see the Appendix \ref{App:assumptionsmoment} for the specific conditions)}, then 
\begin{eqnarray}
\label{eq:generator}
\E(f( X_{t+\Delta})|X_t=x) &=& \sum_{i=0}^k \frac{\Delta^i}{i!} L^if(x) + \mathcal{O}(\Delta^{k+1})
\end{eqnarray}
where $L$ is the generator of model \eqref{GeneralModel}-\eqref{GeneralGamma}, 
\begin{eqnarray*}
Lf(x)&=& (\partial_x f (x)) b(x) +\frac12 \bigtriangledown^2_{\bar
         \Gamma} f(x),
\end{eqnarray*}
and $L^i f$  means $i$ times iterated application of the generator
{\citep[p. 18, Lemma 1.10]{Mangabook}. In particular, it holds
for $f=x$ or $x^2$ for the three models in Section \ref{sec:examples}.} This yields the first
conditional moment of 
the $j$'th 
component of $X_t$, 
\begin{eqnarray}
\label{eq:EXj}
\E( X_{t+\Delta}^{(j)}|X_t=x) &=& x^{(j)} + \Delta b_j(x) +
                                 \frac{\Delta^2}{2} Lb_j(x)
                              +\mathcal{O} (\Delta^3). 
\end{eqnarray}
 In particular, for model \eqref{GeneralModel} we have 
\begin{eqnarray}
\label{eq:VX}
\E( V_{t+\Delta}|X_t=x) &=& v + \Delta a(x) +
\frac{\Delta^2}{2}  \partial_{x} a(x) \, b(x)
+\frac{\Delta^2}{4} \bigtriangledown^2_{\bar \Gamma} 
a(x) +\mathcal{O} (\Delta^3),\\
\label{eq:UX}
\E( U_{t+\Delta}|X_t=x) &=& u + \Delta A (x)+
\frac{\Delta^2}{2} \partial_{x} A(x) \, b(x)
+\frac{\Delta^2}{4} \bigtriangledown^2_{\bar \Gamma} 
A(x) +\mathcal{O} (\Delta^3).
\end{eqnarray}
Furthermore, 
\begin{align}
\mbox{Var} ( V_{t+\Delta}|&X_t=x)
  = \frac{\Delta^3}{3} \partial_u a \Gamma \Gamma^T (\partial_u a)^T +
                            \mathcal{O} (\Delta^4) \\
\mbox{Var} ( U_{t+\Delta}^j|&X_t=x)
  =\Delta  \sigma^2_j (x) + \label{eq:varU} \\ &
                                 \frac{\Delta^2}{2} \left (
      A_{j} \partial_{u_j} \sigma_j^2 (x) + 2\sigma^2_j
      (x) \partial_{u_j} A_j(x) + \frac 12 \sigma_j^2
      (x) \partial_{u_j^2}^2 \sigma_j^2 (x)
\right )
+\mathcal{O} (\Delta^3) \nonumber 
\end{align}

Note how the order of the variance of the first coordinate is
$\Delta^3$, whereas the mean is of order $\Delta$. This is the cause of the
statistical difficulties of estimating the parameters.

\subsection{Three examples}
\label{sec:examples}

\subsubsection{Harmonic Oscillator}\label{sec:HO}
Harmonic oscillators are common in
nature, and the model is central in classical mechanics. 
Consider the damped harmonic oscillator driven by a white noise
forcing \citep{Pokern2009}, 
\begin{equation}\label{eq:HO}\left\{
\begin{array}{lcl}
dV_t&=&U_t dt\\
dU_t&=&(-D V_t - \gamma U_t)dt +  \sigma dB_t
\end{array}
\right.
\end{equation}
with $\gamma, D, \sigma>0$. Here, $p=1$. 
{The drift function $a$ does not depend on an unknown
  parameter}, which makes parameter estimation 
much easier, and thus 
$\beta =\varphi = (D, \gamma$). For this
linear model
we know the true distribution. The process is an
ergodic 
Ornstein-Uhlenbeck process, i.e., a Gaussian process. Define  
$$X_t = \left ( \begin{array}{c} V_t \\ U_t \\ \end{array} \right ) \
; \quad M = \left ( \begin{array}{cc} 0&1 \\ -D&-\gamma \\ \end{array}
\right ) \ ; \quad {C} = \left ( \begin{array}{c} 0 \\ \sigma
                                      \\ \end{array} \right ). $$
Then 
$$dX_t = M X_t dt + {C} dB_t$$
and the conditional distribution is 
\begin{equation}
\label{eq:HOdistribution1}
(X_{t+\Delta} | X_t=x) \sim \mathcal{N} \left (e^{ \Delta M } x  \, ,
  \, \int_0^\Delta e^{ s M }{C C}^T e^{ s M^T }ds \right ). 
\end{equation}
Let $d= \frac 12 \sqrt{\gamma^2-4D}$, then
\begin{equation}
\label{eq:HOdistribution2}
\E (X_{t+\Delta} | X_t=x)=  e^{-\frac{1}{2}\gamma \Delta}
\left ( \begin{array}{c} 
\left (\cosh \left (d\Delta \right )+
\frac{\gamma}{2d} \sinh \left (d\Delta \right ) \right ) x_1+\left (
\frac{1}{d} \sinh \left (d\Delta \right ) \right ) x_2\\
\left (  -\frac{D}{d} \sinh \left (d\Delta \right )  \right ) x_1 + \left (\cosh \left (d\Delta \right )-
\frac{\gamma}{2d} \sinh \left (d\Delta \right ) \right )
          x_2 \end{array} \right ),
\end{equation}
{where we formally define $\sinh (0)/0 = 0$. Note that $d$ has
to be complex  for the solution
to oscillate, i.e., for negative determinant, which is the case we
consider.}
\begin{figure}[tb!]
\includegraphics[width = 0.99\textwidth]{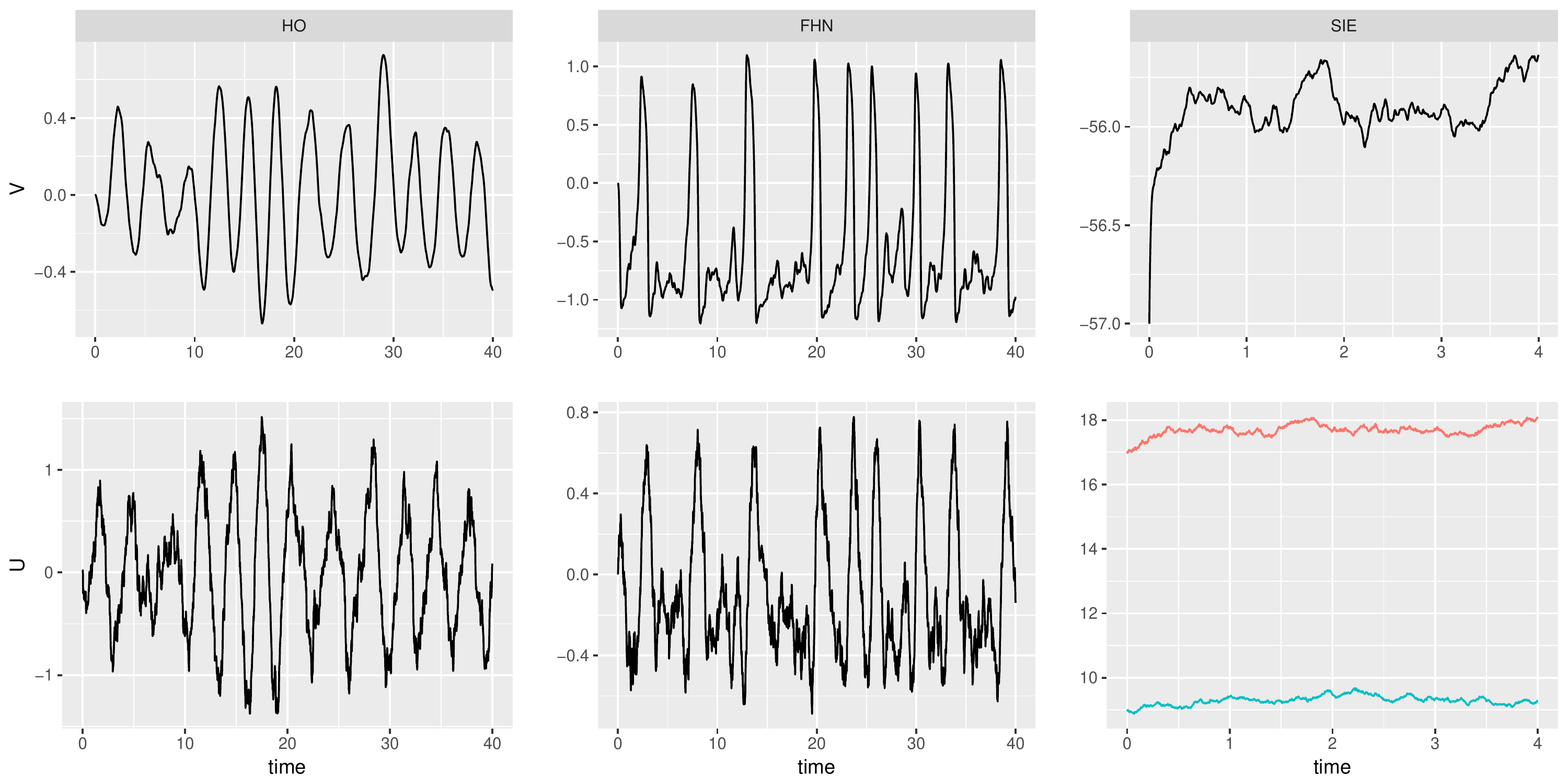}
\caption{\label{fig:traces} Simulated paths of the three example models. Left: Harmonic
  Oscillator. Middle: FitzHugh-Nagumo model. Right: Synaptic
  Inhibition and Excitation model. Upper plots: The smooth coordinate
  $V$. Lower plots: The rough coordinates $U$. {The
    rough paths of the SIE model are excitatory (red) and inhibitory
    (green) conductances.} Parameter values are
  given in Section \ref{sec:simulation}. }
\end{figure}
To compare with the analysis of the other models, we make a Taylor expansion in $\Delta$ up to order 2 obtaining
\begin{equation}
\E (X_{t+\Delta} | X_t=x)=  x + \Delta B_{\mbox{HO}}(x) +\mathcal{O} (\Delta^3) 
\end{equation}
where
\begin{equation}
\label{eq:BHO}
\Delta B_{\mbox{HO}}(x)=  
\Delta \left ( \begin{array}{c} x_2-(Dx_1+\gamma x_2) \frac{\Delta}{2} \\
-(Dx_1+\gamma x_2) + (\gamma(D x_1+\gamma x_2)-Dx_2)
\frac{\Delta}{2}\end{array} \right ).
\end{equation}
Furthermore,
\begin{align}
\label{eq:HOdistribution3}
\mbox{Var} (X_{t+\Delta}& | X_t=x)= \frac{\sigma^2}{2\gamma D}\left
  [ \begin{array}{cc} 1&0\\0 &D\\ \end{array} \right ]+ \\
&\frac{\sigma^2 e^{-\gamma
    \Delta}}{4d^2}\left [ \begin{array}{cc} \frac{2}{\gamma}
    \! - \! \frac{d}{D}\sinh\left (2d \Delta \right ) \! -\! \frac{\gamma}{2D}
    \cosh\left (2d \Delta \right ) &\cosh\left (2d\Delta \right ) -1\\
    \cosh\left (2d\Delta \right ) -1 &\frac{2D}{\gamma} \! +\!
    d\sinh\left (2d \Delta \right ) \! -\! \frac{\gamma}{2} \cosh\left
      (2d \Delta \right )\\ \end{array} \right ] \nonumber
\end{align}
with Taylor expansion up to order 3 in $\Delta$
\begin{align}
\mbox{Var} (X_{t+\Delta}& | X_t=x)= \sigma^2\left
  [ \begin{array}{cc}
    \frac13\Delta^3&\frac12\Delta^2-\frac12\Delta^3\gamma
    \\\frac12\Delta^2-\frac12\Delta^3\gamma
     &\Delta -\gamma \Delta^2+\frac13\Delta^3(2\gamma^2-D)
    \\ \end{array} \right ] +\mathcal{O}(\Delta^4)
\end{align}
where we need a higher order for the variance for later convergence
results, since otherwise the variance of the first coordinate is
zero. 

The invariant distribution is Gaussian, 
$$X_{\infty} \sim \mathcal{N} \left ( 0 , \frac{\sigma^2}{2\gamma D}\left [ \begin{array}{cc} 1&0\\0 &D\\ \end{array} \right ] \right ).$$
The solution of this system has thus moments of any order. An example
path can be found in Figure \ref{fig:traces}.

\subsubsection{FitzHugh-Nagumo}
A prototype of a model of a spiking neuron is the FitzHugh-Nagumo model,
which is a minimal representation of more realistic neuron models, such as
the Hodgkin-Huxley model, modelling the neuronal firing mechanisms 
\citep{FitzHugh1961,Nagumo1962,HodgkinHuxley1952}.

Consider the stochastic hypoelliptic FitzHugh-Nagumo model, defined as
the solution to the system
\begin{equation}\label{eq:FHN}
\left\{
\begin{array}{ccl}
dV_t &=& \frac1\varepsilon (V_t -V_t^3-U_t+s)dt,\\
dU_t&=& \left(\gamma V_t-U_t+{\alpha} \right)dt   +  \sigma dB_t,
\end{array}
\right.
\end{equation}
where the variable $V_t$ represents the membrane potential of
a neuron at time $t$, $U_t$ is a recovery variable, which could represent channel
kinetics{, and} $p=1$. 

Parameter $s$ is the magnitude of the stimulus current.  When only 
$\Vt$ is observed, $s$ is not
identifiable \citep{Jensen2012}. Often $s$
represents injected current and is  
thus controlled in a given experiment, and it is therefore
reasonable to assume it known, so that $\psi=\varepsilon$. Thus, parameters to
be estimated are $\sigma, \psi=(\varepsilon)$ and 
$\varphi=(\gamma, {\alpha})$. 

The distribution of $X_t=(\Vt,\Ut)^T$ is unknown, but moments can be
approximated by using \eqref{eq:generator}, 
where the generator of model \eqref{eq:FHN} is
\begin{eqnarray*}
Lf(x)&=& \frac{1}{\varepsilon}(x_1 -x_1^3-x_2+s) \frac{\partial f}{dx_1} +
(\gamma x_1-x_2+{\alpha}) \frac{\partial f}{dx_2}  +\frac12 \sigma^2 \frac{\partial^2 f}{dx_2^2}.
\end{eqnarray*}
We obtain
\begin{eqnarray*}
\E (X_{t+\Delta} | X_t=x)=  x + \Delta B_{\mbox{FHN}}(x) +\mathcal{O} (\Delta^3) 
\end{eqnarray*}
where
\begin{align}
\label{eq:BFHN}
\Delta &B_{\mbox{FHN}}(x)= \\
&\Delta \left(\begin{array}{c}
\frac1\varepsilon(x_1 - x_1^3 - x_2  +  s) 
+ \displaystyle{\frac{\Delta}{2}}\frac1{\varepsilon}\left
  (\frac1{\varepsilon}(1 - 3x_1^2)(x_1 - x_1^3 - x_2 
  +  s) - (\gamma x_1 - x_2 - {\alpha})
\right )\\
(\gamma x_1 - x_2  + {\alpha})
+\displaystyle{\frac{\Delta}{2}}\left (\frac\gamma\varepsilon(x_1
  - x_1^3 - x_2 + s)-(\gamma x_1 - x_2  +
  {\alpha})\right )
\end{array}\! \! \right)  \nonumber 
\end{align}
and
\begin{align}
\mbox{Var} (X_{t+\Delta}& | X_t=x)= \sigma^2\left
  [ \begin{array}{cc}
    \frac13\Delta^3\frac1{\varepsilon^2}{+\mathcal{O}(\Delta^4)}&-\frac12\Delta^2 \frac{1}{\varepsilon}{+\mathcal{O}(\Delta^3)}\vspace{2mm}
    \\ 
-\frac12\Delta^2 \frac{1}{\varepsilon}{+\mathcal{O}(\Delta^3)} \hspace{2mm}
     &\Delta -\Delta^2 {+\mathcal{O}(\Delta^3)}
    \\ \end{array} \right ].
\end{align}
An example
path can be found in Figure \ref{fig:traces}.

\subsubsection{Synaptic-conductance model}

A neuron, which reliably can be characterized as a single electrical
compartment, and which receives excitatory and inhibitory synaptic bombardment, has a voltage
dynamics across the membrane that can be described by this
conductance-based model with diffusion synaptic input \citep{DayanAbbottBook,BergDitlevsen2013} 
\begin{equation}\label{eq:Conduc}
\left\{
\begin{array}{ccl}
CdV_t &=& (-G_L (V_t-V_L) - G_{E,t} (V_t-V_E) - G_{I,t} (V_t-V_I) + I_{inj} )dt\\[1mm]
dG_{E,t} &=& -\frac{1}{\tau_E}(G_{E,t}-\bar g_E)dt + \sigma_E\sqrt{G_{E,t}}dB_{E,t}\\[1mm]
dG_{I,t} &=& -\frac{1}{\tau_I}(G_{I,t}-\bar g_I)dt + \sigma_I \sqrt{G_{I,t}}dB_{I,t}
\end{array}
\right.
\end{equation}
 where $C$ is the total capacitance, $G_L$, $G_E$ and $G_I$ are the
 leak, excitation, and inhibition conductances, $V_L$, $V_E$ and $V_I$
 are their respective reversal potentials, and $I_{inj}$ is the injected
 current.  The conductances $G_{{E},t}$ and $G_{{I},t}$ are assumed to be stochastic
 functions of time, where $(B_{E,t})$ and $(B_{I,t})$ are two
 independent Brownian motions. The square roots in the diffusion
 coefficient ensures that the conductances stay positive. Parameters $\tau_E, \tau_I$ are time
 constants, $\bar g_E, \bar g_I$ the mean conductances, and $\sigma_E,
 \sigma_I$ the diffusion coefficients, scaling the variability of these two
 processes.  Here, $U_t=(G_{{E},t}, G_{{I},t})^T$ and $p=2$. We assume the
 capacitance and the reversal
 potentials {known}, which are easily 
 determined in independent experiments \citep{BergDitlevsen2013}, as
 well as $I_{inj}$, which is controlled by the experimenter. Thus,
 {the drift function $a$ does not depend on an unknown 
  parameter}, $\varphi=(\bar g_E, \bar g_I, \tau_E,
 \tau_I)$, and $\sigma=(\sigma_E,  \sigma_I)$.  

{The distribution of $\Vt$ is also unknown for this
model, whereas $\Ut$ are independent square root processes (also
called CIR processes), which have transition densities following
non-central chi-square distributions. However, for illustration of the
methodology, we
will approximate moments} by using the generator of model \eqref{eq:Conduc}, 
\begin{eqnarray*}
Lf(x)&=& \frac{1}{C}(-G_L (x_1-V_L) - x_2 (x_1-V_E) - x_3 (x_1-V_I) +
         I_{inj} ) \frac{\partial f}{dx_1} \\
&&-\frac{1}{\tau_E}(x_2-\bar g_E) \frac{\partial f}{dx_2}
   -\frac{1}{\tau_I}(x_3-\bar g_I) \frac{\partial f}{dx_3 }
   +\frac12 \sigma_E^2 x_2\frac{\partial^2 f}{dx_2^2} +\frac12
   \sigma_I^2 x_3\frac{\partial^2 f}{dx_3^2}  
\end{eqnarray*}
and equation \eqref{eq:generator}. 
We obtain
\begin{align}
\E &(X_{t+\Delta})|X_t=x)= x+ \Delta B_{\mbox{SIE}}(x) +\mathcal{O} (\Delta^3) 
\end{align}
where
\begin{equation}
\label{eq:BSIE}
\Delta B_{\mbox{SIE}}(x)\! =\! \Delta \! \left(\begin{array}{c}
\! \! \! b_1(x)
 \displaystyle{-\frac{\Delta}{2C}} \! \left ( b_1(x)
                                        (G_L\! +\! x_2 \! + \! x_3) \! +
       \!  b_2(x) (x_1\! -\! V_E) +   b_3(x) (x_1 \! -\! V_I) 
\right )\\
b_2(x)-\displaystyle{\frac{\Delta}{2}}\left (
        b_2(x) \frac{1}{\tau_E} \right )\\
b_3(x) -\displaystyle{\frac{\Delta}{2}}\left (
        b_3(x) \frac{1}{\tau_I} \right )
\end{array}\! \! \right) \! 
\end{equation}
and
{\begin{align}
&\mbox{Var} (X_{t+\Delta} | X_t=x)=\\
& \left
  [ \begin{array}{ccc}
\frac{\Delta^3}{3C^2}((x_1\! -\! V_E)^2\sigma_E^2x_2\! +\! (x_1\! -\! V_I)^2\sigma_I^2x_3) {+\mathcal{O}(\Delta^4)}   & -\frac{\Delta^2}{2C} \sigma_E^2x_2( x_1-V_E) {+\mathcal{O}(\Delta^3)}& -\frac{\Delta^2}{2C} \sigma_I^2x_3( x_1-V_I) {+\mathcal{O}(\Delta^3)}\vspace{2mm}
    \\ 
-\frac{\Delta^2}{2C} \sigma_E^2x_2( x_1-V_E) {+\mathcal{O}(\Delta^3)}
                                                                                  &
\Delta\sigma_E^2 x_2 {+\mathcal{O}(\Delta^2)}
&0 \vspace{2mm}
    \\ 
-\frac{\Delta^2}{2C} \sigma_I^2x_3( x_1-V_I)
      {+\mathcal{O}(\Delta^3)}&0&
\Delta\sigma_I^2 x_3 {+\mathcal{O}(\Delta^2)}
\end{array} \right ]. \nonumber
\end{align}}
An example
path can be found in Figure \ref{fig:traces}. The red path is the
excitatory conductance, the green path is the inhibitory conductance.

\section{Discretization scheme}
\label{sec:discretization}

The transition density for model
\eqref{GeneralModel} is generally unknown, and a possible approximation to the
likelihood function is the likelihood for some approximating scheme of
the discretized 
process $X_{0:n}$. We will write $\tilde X_i$ for
the approximated process, or $\tilde V_i$ and $ \tilde
U_i$ where relevant. 

The most commonly applied scheme to approximate the likelihood in
SDEs, especially for high-frequency data, is the Euler-Maruyama
approximation of model  (\ref{GeneralModel}), which leads
to a discretized model defined as follows 
\begin{eqnarray}\label{eq:discretize_ML_2}
\tilde V_{i+1} &=& \tilde \Vi+\Delta a(\tilde \Vi,\tilde \Ui),\\
\tilde U_{i+1}&=&\tilde \Ui + \Delta A(\tilde \Vi,\tilde \Ui) + \Gamma(\tilde \Vi,\tilde \Ui)\eta_i,\nonumber
\end{eqnarray}
where   $(\eta_i)$ are centered Gaussian vectors with variance
$\Delta I_p$. 
 Thus, the transition density of the approximate discretized scheme is
a degenerate Gaussian distribution, since there is no stochastic term
on the first coordinate. The same  
 happens for the Milstein-scheme with strong order of convergence
 equal to 1.

\subsection{Discretization with 1.5 scheme}

 We propose to use a higher order scheme, namely the 1.5 strong order
 scheme \citep{KloedenPlatenBook}, using the hypoellipticity of (\ref{GeneralModel}) to
 propagate the noise into the first coordinate. For a diagonal diffusion matrix as in
 \eqref{GeneralGamma} the scheme is as follows, where 
 for readability we have suppressed the dependence on $(\Vi,\Ui)$,
\begin{eqnarray}\label{eq:discretize_strong}
\tilde V_{i+1} &=& \tilde \Vi+\Delta a +
\frac{\Delta^2}{2} \partial_{x} a \, b
+\frac{\Delta^2}{4} \bigtriangledown^2_{\bar \Gamma} 
a  + \partial_{u} a \, \Gamma \xi_i \label{eq:discretize_strongV}\\
\tilde U_{i+1} &=&\tilde \Ui + \Delta A + 
\frac{\Delta^2}{2} \partial_{x} A \, b+\frac{\Delta^2}{4} 
\bigtriangledown^2_{\bar \Gamma}  A  + \Gamma \eta_i +\partial_{u} A \, \Gamma  \xi_i 
\nonumber \\
&& \label{eq:discretize_strongU}
+\frac12 \partial_{u} \bar \Gamma  \, \Gamma (\eta_i^{*2}
-{\Delta}\mathbf{1}_p)+\partial_{u} \bar \Gamma  \, A (\Delta \eta_i -\xi_i) +
\frac12 \bigtriangledown^2_{\bar \Gamma }  \bar \Gamma  (\Delta \eta_i -\xi_i)
\nonumber \\
&&+
\frac12 {\left ( \left (  \partial_{u}  \bar \Gamma
\right )^2 \Gamma + \bigtriangledown^2_{\bar \Gamma }  \bar \Gamma\right )} (\frac13 \eta_i \eta_i^T - \Delta I_p)\eta_i
\end{eqnarray}
where $(\eta_i)$ are centered Gaussian vectors with variance
$\Delta I_p$, $(\xi_i)$ are centered Gaussian vectors with variance
$\Delta^3/3 I_p$, Cov$(\eta_i,\xi_i)= \Delta^2/2 I_p$ and
Cov$(\eta_i,\xi_j)= 0$ for $i \neq j$. Furthermore, $\eta_i^{*2}$
denotes the vector with the squared entries of $\eta_i$. Notice how noise of order
$\Delta^{3/2}$ is now propagated into the first equation, since the
last term on the right hand side of \eqref{eq:discretize_strong} is
non-zero if condition {\bf (C1)} is fulfilled. If $\Gamma$
is independent of the process (additive noise) then the last two lines
in \eqref{eq:discretize_strongU} are zero. 

To simplify the notation later on, we rewrite equations
(\ref{eq:discretize_strongV})-(\ref{eq:discretize_strongU}) as 
\begin{equation}\label{eq:discretize_strongSimple}
\left(\begin{array}{c} \tilde V_{i+1}\\\tilde U_{i+1}\end{array}\right) = 
\left(\begin{array}{c}\tilde \Vi\\\tilde \Ui\end{array}\right) + \Delta B(\tilde \Vi, \tilde \Ui) +
\varepsilon_i, \quad \varepsilon_i\sim \mathcal{N}_{p+1}(0, \Sigma (\tilde \Vi, \tilde \Ui))
\end{equation} 
where $\Delta B(v,u)_j = \Delta b_j +
\frac{\Delta^2}{2} \partial_{x} b_j \, b
+\frac{\Delta^2}{4} \bigtriangledown^2_{\bar \Gamma} 
b_j $ is the scheme for the drift and $\Sigma (v,u)$
is the variance matrix of the scheme.  Up to leading order, the
variance matrix is given by
\begin{equation}
\label{Sigma}
\Sigma (v,u) = \left(\begin{array}{cc} 
\partial_{u} a \, \Gamma \Gamma^T (\partial_{u} a)^T
                       \frac{\Delta^3}{3}& \partial_{u} a \, \Gamma
                                           \Gamma^T
                                           \frac{\Delta^2}{2}\\ 
\Gamma \Gamma^T (\partial_{u} a)^T
                       \frac{\Delta^2}{2}& \Gamma \Gamma^T \Delta\\
\end{array}\right) .
\end{equation} 
Since the mean term
coincides with the true mean up to order $\Delta^2$, see eqs. \eqref{eq:VX}
and \eqref{eq:UX}, the functions ${\Delta}B(v,u)$ for models \eqref{eq:HO},
\eqref{eq:FHN},  and \eqref{eq:Conduc} are given in \eqref{eq:BHO}, \eqref{eq:BFHN}
and \eqref{eq:BSIE}, respectively. The variance matrix
$\Sigma (\tilde V_{i},\tilde U_{i})$ of the above scheme for the three models  \eqref{eq:HO}, 
\eqref{eq:FHN},  and \eqref{eq:Conduc} are
\begin{align}  
\label{eq:sigmaHO}
&\, \Sigma_{\mbox{HO}} \, \, = \sigma^2\left(\begin{array}{cc}
\frac13 \Delta^3&\frac12 \Delta^2-\frac13 \Delta^3 \gamma\\
\frac12 \Delta^2-\frac13 \Delta^3 \gamma&\Delta-\Delta^2 \gamma+\frac13 \Delta^3 \gamma^2\\
\end{array}\right),\\
&\Sigma_{\mbox{FHN}} = \sigma^2\left(\begin{array}{cc}
\frac13 \Delta^3 \varepsilon^{-2}&\left ( -\frac12 \Delta^2+\frac13
  \Delta^3 \right ) \varepsilon^{-1}\\
\left ( -\frac12 \Delta^2+\frac13
  \Delta^3 \right ) \varepsilon^{-1} &\Delta-\Delta^2 +\frac13 \Delta^3 \\
\end{array}\! \! \right),\\
&\, \Sigma_{\mbox{SIE}} (\tilde V_i, \tilde U_i)\, \, = \\
&\left( \begin{array}{ccc}
\! \! \! \! \frac{\Delta^3}{3}\!  \left ( {(\tilde\Vi \!
          -\!V_E)^2}\sigma^2_E \tilde G_{E,i}  \! + \! { (\tilde\Vi \! -\!V_I)^2}\sigma^2_I \tilde G_{I,i} 
  \right )& -  \sigma^2_E { (\tilde\Vi \! -\! V_E)} \tilde G_{E,i} \! \left ( \frac{\Delta^2}{2} \! \! +\!
    \frac{\Delta^3}{6\tau_E}\right )& -  \sigma^2_I { (\tilde\Vi \! -\!V_I)} \tilde G_{I,i} \! \left ( \frac{\Delta^2}{2} \! \! +\!
    \frac{\Delta^3}{6\tau_I}\right )\\
 -  \sigma^2_E { (\tilde\Vi-V_E)} \tilde G_{E,i} \! \left ( \frac{\Delta^2}{2} \! \! +\!
    \frac{\Delta^3}{6\tau_E}\right ) & \sigma^2_E \tilde G_{E,i} \left ( \Delta-\frac{\Delta^2}{2\tau_E}+\frac{\Delta^3}{12\tau_E^2}\right )&0\\
-  \sigma^2_I { (\tilde\Vi-V_I)} \tilde G_{I,i} \! \left ( \frac{\Delta^2}{2} \! \! +\!
    \frac{\Delta^3}{6\tau_I}\right )&0&\sigma^2_I \tilde G_{I,i} \left (
                                         \Delta \! - \!
                                         \frac{\Delta^2}{2\tau_I}\! +
                                         \! \frac{\Delta^3}{12\tau_I^2}\right )\\
\end{array}\right). \nonumber
\end{align}
For comparison, we recall the variance matrix for the HO model
suggested by \cite{Pokern2009},
\begin{align}  
&\, \Sigma_{\mbox{HO, Pokern}} \, \, = \sigma^2\left(\begin{array}{cc}
\frac13 \Delta^3&\frac12 \Delta^2\\
\frac12 \Delta^2&\Delta\\
\end{array}\right),
\end{align}  
which coincides with \eqref{eq:sigmaHO} up to lowest order at each
matrix entry. Furthermore, it coincides with \eqref{Sigma} when $p=1,
a(v,u)=u$ and $ \Gamma (v,u) = \sigma$.

\subsection{{Remarks on the convergence of the scheme}}

The scheme
\eqref{eq:discretize_strong}--\eqref{eq:discretize_strongU} has a 
strong order 1.5 and a weak order 2 convergence \citep{KloedenPlatenBook}. The following bounds
follow by comparing eqs. \eqref{eq:generator}--\eqref{eq:varU} with
eqs. \eqref{eq:discretize_strong}--\eqref{eq:discretize_strongU}. These
bounds are needed later to prove
consistency.

\begin{prop}\label{prop:order_scheme}[{Moment bounds}]
\begin{align*}
\E (V_{i+1}-V_i -\Delta B(X_i)_1 | X_i { =x}) &= \mathcal{O} (\Delta^3) \\
\E (U_{i+1}-U_i -\Delta B(X_i)_{(-1)} | X_i { =x}) &= \mathcal{O} (\Delta^3) \\
\E ((V_{i+1}-V_i -\Delta B(X_i)_1 )^2| X_i { =x}) &= \frac{\Delta^3}{3} \partial_u a \Gamma \Gamma^T (\partial_u a)^T +
                            \mathcal{O} (\Delta^4) \\
\E ((U_{i+1}\! -\!  U_i \! -\! \Delta B(X_i)_{(-1)} ) (U_{i+1}\! -\!  U_i \! -\! \Delta B(X_i)_{(-1)} )^T| X_i { =x}) &= \Delta \Gamma
                                                        \Gamma^T +\mathcal{O}
(\Delta^2) \\
\E ((V_{i+1} -  V_i -\Delta B(X_i)_1 )^4| X_i { =x}) &= \mathcal{O} (\Delta^4) \\
\E (((U_{i+1} \! -\!  U_i \! -\! \Delta B(X_i)_{(-1)} ) (U_{i+1}\! -\!  U_i \! -\! \Delta B(X_i)_{(-1)} )^T)^2| X_i { =x}) &= \mathcal{O}
(\Delta^2) \\
\end{align*}
{where $B(X_i)_{(-1)}$ denotes the vector $B(X_i)$ with
  the first coordinate omitted. }
\end{prop}
 
{Note that the expected value of the difference between the true
  drift and the approximating drift $\Delta B$
is of order $\Delta^3$, 
because of the higher order scheme. This is necessary for the later
convergence results in Propositions \ref{prop:consistency2} and
\ref{prop:consistency1}, in particular, 
the technical lemmas of Section \ref{sec:technicallemmas}.}

{Another useful convergence result  is the convergence of the transition density of the scheme to the exact transition density, as  proved in the elliptic case by \cite{Bally1996} under smooth conditions on the drift functions and diffusion coefficients. 
Unfortunately, this result is much more difficult to obtain for a hypoelliptic SDE
such as system (\ref{GeneralModel}). This is beyond the scope of this paper.}

\section{Complete observations}\label{sec:bothobserved}

In this Section we investigate parameter estimation when all
coordinates are discretely observed. Later, we extend to the situation
where only the first coordinate is observed.

\subsection{Contrast estimator}

The goal is to estimate the parameter $\theta=(\psi, \varphi, \sigma)$
by maximum likelihood of the approximate model, with complete
likelihood  
\begin{equation}\label{eq:completeLikelihood_approx}
\pDelta(\Vton,\Uton;\theta) =    p(\Vz,\Uz;\theta)\prod_{i=1}^n \pDelta(\Vi, \Ui|\Vim, \Uim;\theta),
\end{equation}
{where $p(\Vz,\Uz;\theta)$ is the density of the
  initial value of the process. The contribution from this single data point is
  negligible for relevant sample sizes, and we will simply assume it
  degenerate in the observed value $(\Vz,\Uz)$. }
{This likelihood} corresponds to a pseudo-likelihood for the exact diffusion, with
exact complete likelihood given in \eqref{eq:completeLikelihood}.  
The estimator is then the minimizer of minus 2 times the log complete likelihood:
\begin{align}\label{eq:pseudoLike}
\arg\min_{\theta}& \sum_{i=0}^{n-1}\left( (X_{i+1}\! -\!  X_i-\Delta B(X_i;
                   \theta))^T \, \Sigma_i^{-1}(X_{i+1}\! -\! X_{i}-\Delta
                   B(X_i; \theta))  
+  \log \det (\Sigma_i)\right)
 \end{align}
This {criterion} is ill behaved because the system is
hypoelliptic, so the order of the variance for $V$ is $\Delta^3$ and 
{for $U$ it is $\Delta$}. Therefore, we propose to separate the estimation of 
parameter $\psi$ of the first coordinate from parameters $(\phi,
\sigma)$ of the second coordinate.  

We thus introduce two new contrasts and their corresponding estimators.
\begin{definition}
The estimator of the parameters of the first coordinate is given by
\begin{eqnarray}\label{eq:hatpsi}
\hat \psi_n &= & \arg\min_{\psi} \left( \frac{3}{\Delta^3}
                 \sum_{i=0}^{n-1} \frac{(V_{i+1}-V_i - \Delta B(X_i;
                 \theta)_1)^2}{(\partial_u a(X_i;\psi))
                 \Gamma\Gamma^T(X_i; \sigma)(\partial_u
                 a(X_i;\psi))^T} \right.\\ 
&&\left. + \sum_{i=0}^{n-1} \log ((\partial_u a(X_i;\psi))
   \Gamma\Gamma^T(X_i; \sigma)(\partial_u a(X_i;\psi))^T)\right)
   \nonumber 
\end{eqnarray}
where the parameters $\varphi$ and $\sigma$ are {fixed}. 

The estimator of the parameters of the second coordinate is given by
\begin{eqnarray}
(\hat \varphi_n, \hat \sigma_n^2 ) &=   &\arg\min_{\varphi, \sigma^2}
   \left( \sum_{i=0}^{n-1} \log (\det(\Gamma\Gamma^T(X_i;
   \sigma))\right.\label{eq:hatphi} 
  \\
&+&\left.  \sum_{i=0}^{n-1}  (U_{i+1}\! -\! U_i \! - \Delta B(X_i;
   \theta)_{(-1)})^T \left(\Delta \Gamma\Gamma^T\! (X_i; 
   \sigma)\right)^{-1} \! 
    (U_{i+1}\! -\! U_i \! -\Delta  B(X_i; \theta)_{(-1)})\right)   \nonumber 
\end{eqnarray}
where the parameter $\psi$ is {fixed}. 
\end{definition}
The first contrast corresponds to the 
pseudo-likelihood of the marginal distribution of the first
coordinate. The second contrast is a simplification of the
pseudo-likelihood of the marginal of the coordinates with direct
noise: the variance appearing 
in the pseudo-likelihood is $\Delta \Gamma\Gamma^T(X_i, \sigma)(1+
o(\Delta))$ and is simplified to $\Delta \Gamma\Gamma^T(X_i, \sigma)$
in the contrast (\ref{eq:hatphi}), since the variance is dominated by 
the lowest order term. 
The contrasts (\ref{eq:hatpsi}) and  (\ref{eq:hatphi}) require the
other parameters to be fixed. {To estimate the complete
  parameter vector, the parameters are initialized and then the
  optimization procedure iterates between the two estimators (\ref{eq:hatpsi})
  and (\ref{eq:hatphi}).} The numerical 
optimization of the criteria is not sensitive to those fixed values
{since} 
they appear in higher order term{s}.

\subsection{Theoretical properties of the contrast estimators}
We start by proving the consistency of the contrast
estimators. The asymptotics are in number of observations $n$ and
length of time
step between observations $\Delta_n$, where we have introduced an
index $n$ to clarify the relevant asymptotics.

\begin{prop} \label{prop:consistency2} {Assume the
    drift function $a$ can be decomposed 
as either: $a(x;\psi) = a_v(v, \psi) + a_{u}(u)$ or $a(x;\psi) = a_v(v) + \psi a_u(x)$.}
Denote by $\psi_0$ the true value of the parameter, and assume
$(\varphi, \sigma^2)$ known. 
If $\Delta_n \rightarrow 0$ and $n\Delta_n\rightarrow \infty$  then 
$$\hat \psi_n \stackrel {P}{\rightarrow} \psi_0 .$$
\end{prop}
\begin{prop} \label{prop:consistency1} 
Denote by $(\varphi_0, \sigma^2_0)$ the true values of the parameters, and assume
$\psi$ known. 
If $\Delta_n \rightarrow 0$ and $n\Delta_n\rightarrow \infty$  then 
$$(\hat \varphi_n , \hat \sigma^2_n) \stackrel {P}{\rightarrow} (\varphi_0, \sigma^2_0). $$
\end{prop}
The proofs are given in {Supplementary Material,
  Section \ref{append_consistency}. In the numerical examples, the
  parameters are estimated and not fixed to their true values.}
 
 The convergence conditions are standard: the length of the
 observation interval has to increase for consistency of drift
 parameters. For consistency of the variance parameter, it can be
 proven that only $\Delta_n \rightarrow 0$ and $n\rightarrow \infty$
 are needed, but we will not pursue that here.

The estimators are asymptotically
normal. We
prove the result for $(\hat \varphi_n, \hat \sigma^2_n)$ and give some
partial proofs for $\hat \psi_n$.  

\begin{theorem} \label{prop:normality}
Let $\nu(\cdot)$ denote the stationary density of model
\eqref{GeneralModel}. 
 If $\Delta_n \rightarrow 0$,  $n\Delta_n\rightarrow \infty$ and $n\Delta_n^2\rightarrow 0$, then 
\begin{eqnarray*}
\sqrt{n\Delta_n} (\hat \varphi_n - \varphi_0)&\stackrel
                                               {\mathcal{D}}{\rightarrow}
  & \mathcal{N}\left(0, \left(\nu\left( (\partial_\varphi A(\cdot,
    \varphi_0))^T(\Gamma\Gamma^T(\cdot, \sigma_0))^{-1}(\partial_\varphi
    A(\cdot, \varphi_0))\right)\right)^{-1}\right)\\ 
\sqrt{n} (\hat \sigma_n - \sigma_0)&\stackrel
                                     {\mathcal{D}}{\rightarrow}  &
                                                                   \mathcal{N}\left(0,
                                                                   2\left(\nu\left(
                                                                   (\partial_\sigma
                                                                   \Gamma\Gamma^T(\cdot,
                                                                   \varphi_0))^T(\Gamma\Gamma^T(\cdot,
                                                                   \sigma_0))^{-1}(\partial_\sigma
                                                                   \Gamma\Gamma^T(\cdot,
                                                                   \varphi_0))\right)\right)^{2}\right)   
\end{eqnarray*}
{where $\nu(f(\cdot))=\int f(x) d \nu(x)$.}
\end{theorem}
{For the estimator of the parameters of the smooth
  coordinate, the rate of convergence is faster.}
\begin{theorem} \label{prop:normality2}
Let $\nu(\cdot)$ denote the stationary density of model
\eqref{GeneralModel}. Assume the drift function $a$ can be decomposed
as either: $a(x;\psi) = a_v(v, \psi) + a_{u}(u)$ or $a(x;\psi) = a_v(v) + \psi a_u(x)$.
If $\Delta_n \rightarrow 0$,  $n\Delta_n\rightarrow \infty$ and $n\Delta_n^2\rightarrow 0$, then 
\begin{eqnarray*}
\sqrt{\frac{n}{\Delta_n}} (\hat \psi_n - \psi_0)&\stackrel
                                               {\mathcal{D}}{\rightarrow}
  & \mathcal{N}\left(0, \frac13 \left(\nu\left( (\partial_\psi a(\cdot,
    \psi_0))^T(\partial_ua(\cdot,\psi_0)\Gamma\Gamma^T(\cdot, \sigma_0) (\partial_ua(\cdot,\psi_0))^T)^{-1}(\partial_\psi
    a(\cdot, {\psi_0}))\right)\right)^{-1}\right)
\end{eqnarray*}
\end{theorem}

The proofs are given in {Supplementary Material, Section \ref{append_consistency}}.

\section{Partial observations}\label{sec:unobserved}
In this Section we assume that we do not observe the coordinates
$U_t$, which is the most relevant case for applications. The
likelihood to maximize is therefore not the complete approximate
likelihood, but the approximate likelihood $\pDelta(\Vton; \theta)$
defined as the integral of the complete approximate likelihood
(\ref{eq:completeLikelihood_approx}) with respect to the hidden
components.  
\begin{equation}\label{eq:likelihood_approx} 
\pDelta(\Vton;\theta) =    \int \prod_{i=1}^n \pDelta(X_i|X_{i-1};\theta) d\Uton.
\end{equation}
It corresponds to a discretization of the exact likelihood (\ref{eq:Likelihood}). 

The multiple integrals of equation (\ref{eq:likelihood_approx}) are
difficult to handle and it is not possible to maximize the
pseudo-likelihood directly.  As explained in Section \ref{sec:bothobserved}, it is easier to maximize the complete approximate likelihood, after imputing the hidden coordinates. 

For models where $a(v,u)=a_v(v) + a_u(v)u$ for some functions {$a_v$ and $a_u$}
that do not depend on the parameter, such as in the
HO model, the imputation is intuitive: the unobserved coordinate
$U_t$ can be approximated by the differences of the observed
coordinate $V_t$, $U_{i} \approx( (V_{i+1}-V_i)/\Delta-a_v(V_i))/a_u(V_i)$. However, this
induces a bias in the estimation of $\sigma$ \citep[see][for more
details]{Samson2012}, and is moreover only applicable for drift
functions of the observed coordinate such that $u$ can be isolated. We
will take advantage of that when initializing the estimation
algorithm in Section \ref{sec:initialization}.

In this paper we propose to use a particle filter, also known as
Sequential Monte Carlo (SMC), to impute the hidden coordinates. Then,
this imputed path is plugged into a stochastic  SAEM algorithm 
\citep{Delyon1999}, as done in \cite{Ditlevsen2014} for the elliptic case. The SMC proposed by
\cite{Ditlevsen2014} allows to filter a hidden coordinate that is not
autonomous in the sense that the equation for $U_t$ depends on the
first coordinate $V_t$. Here, we extend the algorithm to the case of
$p$ hidden coordinates, to deal with a $p+1$-dimensional SDE.  

More precisely, the observable vector $\Vton$ is then part of a so-called complete vector $(\Vton, \Uton)$, where $\Uton$ has to be
imputed. At each iteration of the SAEM algorithm, the unobserved data
are filtered  under the smoothing distribution 
$\pDelta (\Uton\;|\Vton;  \theta)$  with an SMC. Then the parameters are updated using the pseudo-likelihood proposed in Section \ref{sec:bothobserved}. 
Details on the filtering are given in  Section
\ref{sec:filtering}, and the SAEM algorithm is presented in Section 
\ref{sec:SAEM}.

\subsection{Particle filter}\label{sec:filtering}
 The SMC proposed in \cite{Ditlevsen2014} is designed for a $p=1$-dimensional
 hidden coordinate. Here we extend to the general
 case. For notational simplicity, $\theta$ is   omitted in the rest of
 this Section.

The SMC 
algorithm provides $K$ particles 
$(\Utonk)_{k=1, \dots, K}$ and weights $(\Weighttonk)_{k=1, \dots, K}$ such that the empirical measure 
$\Psi^K_{n} = \sum_{k=1}^K \Weightn(\Uton^{(k)}) \mathbf{1}_{\Uton^{(k)}} $
approximates the conditional smoothing distribution
$\pDelta(\Uton|\Vton)$ \citep[]{Doucet2001}.   
The SMC method relies on  proposal distributions $q(\Ui | \Vi,\Vim,
\Uim)$ to sample the particles from these
distributions. We write $\Vtoi = (\V_0, \ldots , \Vi)$ and likewise
for  $\Utoi$.

\bigskip

\begin{algo}[\textbf{SMC  algorithm}] \label{SMCalgo}
$\;$ 
\begin{itemize}
\item \emph{At time $i=0$: } $\forall\, k = 1,\ldots, K$
\begin{enumerate}
\item sample $\Uz^{(k)}$  from $p(\Uz|\Vz)$
\item compute and normalize the weights:\\
$\weightz\left(\Uzk\right) = {1}$,  $\quad \Weightz \left(\Uzk\right) = \frac{\weightz
 \left(\Uzk\right) }{\sum_{k=1}^{K} \weightz
 \left(\Uzk\right) } $
\end{enumerate}
\item \emph{At time $i=1,\ldots, n$: } $\forall\, k = 1,\ldots, K$
\begin{enumerate}
\item sample indices  $A_{i-1}^{(k)}\sim r(\cdot |
  \Weightim(\Utoim^{(1)}), \ldots,  \Weightim(\Utoim^{(K)}) )$ {where $r(\cdot)$ denotes the multinomial distribution} and set
  \\
$\Utoim^{'(k)} = \Utoim^{(A_{i-1}^{(k)})}$ 
\item sample $\Uik \sim q \left(\cdot  | \V_{i-1:i}, \U^{'(k)}_{i-1} \right)$ and set $\Utoik = (\Utoim^{'(k)}, \Uik)$ 
\item compute and normalize the weights $\Weighti(\Utoik) = \frac{\weighti \left(\Utoik\right) }{\sum_{k=1}^{K} \weighti \left(\Utoik\right)}$ with \\
{ $\weighti\left(\Utoik \right) = \frac{\pDelta \left(\Vtoi, \Utoik\right)}{\pDelta\left(\Vtoim,\Utoim^{'(k)} \right)q\left(\Uik |  \V_{i-1:i},\Utoim^{'(k)}\right)}
$}
\end{enumerate}
\end{itemize}
\end{algo}

Natural choices for the proposal $q$ are either the
transition density $q(\Ui | \V_{i-1:i},\Uim) = \pDelta(\Ui| \Vim,\Uim)$ or the conditional distribution $q(\Ui|\V_{i-1:i},\Uim) = \pDelta(\Ui|
\V_{i-1:i},\Uim)$, following \cite{Ditlevsen2014}. The two choices are not equivalent in the hypoelliptic case because the covariance matrix of the approximate scheme is not diagonal.
The conditional distribution gives better results in practice and is
used in the simulations. This is due to the extra information provided
by also conditioning on $V_i$.

In the following, we present some asymptotic convergence results on
the SMC algorithm. The assumptions can be found in
{Supplementary Material, Section \ref{App:assumptions}}. 
For a bounded Borel function $f$, 
denote 
$\Psi_{n}^K {(f)} = \sum_{k=1}^K f(\Un^{(k)})
\Weight_{n}(\Uton^{(k)})$, the conditional expectation of $f$
under the empirical measure $\Psi_{n}^K$. We also denote $\pi_{n,\Delta} {(f)} = \EDelta\left(f(\Un)|\Vton\right)$ 
the conditional expectation under the smoothing distribution
$\pDelta(\Uton|\Vton)$ of the approximate model.

\begin{prop}\label{lemma}
Under assumption (SMC3), for any $\varepsilon>0$, and for any bounded
Borel function $f$ on $\mathbb{R}$, there exist constants {$C_{1,\Delta}$ and
$C_{2,\Delta} $ that do not depend on $K$}, such that 
\begin{eqnarray}
\label{lemma1}
\mathbb{P}\left(\left|\Psi_{n}^K {(f)} - \pi_{n,\Delta} {(f)}\right|\geq \varepsilon\right)&\leq& C_{1,\Delta} \exp\left(-K \frac{\varepsilon^2}{C_{2,\Delta} \|f\|^2}\right) 
\end{eqnarray}
where $\| f\|$ is the sup-norm of $f$ and $C_{1,\Delta}$, $C_{2,\Delta}$ are constants detailed in \cite{Ditlevsen2014}.
\end{prop}
The proof is the same as in \cite{Ditlevsen2014}. The hypoellipticity
of the process is not a problem as the filter is applied on the
discretized process where the noise has been propagated to the first
coordinate{, such that the ratio in Algorithm 1, step (c)
  will be well-defined when calculating the weights, since then $p_{\Delta}$
and $q$ are non-degenerate normal densities different from 0}.

\subsection{SAEM}\label{sec:SAEM}
 
The estimation method is based on a stochastic version of the EM
algorithm \citep{Dempster1977}, namely the SAEM algorithm
\citep{Delyon1999} coupled to the SMC algorithm, as already proposed
by \cite{Ditlevsen2014} in the elliptic case. 
To fulfill convergence conditions of the algorithm, we consider the
particular case of a distribution from an exponential
family. {Note that it is the discrete pseudo-likelihood \eqref{eq:completeLikelihood_approx} using the
  strong order 1.5 scheme that needs to fulfill the conditions.} More
precisely, we assume: 
\begin{itemize}
\item[\textbf{(M1)}] The parameter space $\Theta$ is an open subset of
  $\mathbb{R}^p$. The complete {pseudo-}likelihood 
  belongs to a curved exponential family, i.e.,  
$\log \pDelta(\Vton,\Uton;\theta)=  - \psi(\theta) + \left\langle S(\Vton,\Uton),\nu(\theta)\right\rangle$,
where $\psi$ and $\nu$ are two functions of  $\theta$,  $S(\Vton,\Uton)$ is
known as the minimal sufficient statistic of the complete model,
taking its value in a subset ${\mathcal{S}}$ of $\mathbb{R}^d$, and $
\left\langle\cdot, 
\cdot\right\rangle$ is the scalar product on $\mathbb{R}^d$. 
\end{itemize}
The three models considered in this paper satisfy this
assumption. Details of the sufficient statistic $S$ for the HO model are
 given in {the Supplementary Material,} Appendix \ref{sec:suff_stat_HO}. 

Under assumption (M1), introducing a sequence of positive numbers
$(a_m)_{m\in \mathbb{N}}$ decreasing to 
zero, the SAEM-SMC algorithm is defined as follows. 

\newpage

\begin{algo}[SAEM-SMC algorithm]
$\;$ 
\vspace{-0.5em}
\begin{itemize}
\item \emph{Iteration $0$:}  initialization of   $\; \hthetaz$ and set 
  $s_{0}=0$.
\item \emph{Iteration $m\geq 1$:}

\begin{itemize}
   \item[\emph{S-Step:}] simulation of the non-observed data
     $(\Uton^{(m)})$ with SMC targeting the  {smoothing }
     distribution $\pDelta(\Uton|\Vton; \hthetamm)$. 
   \item[\emph{SA-Step:}] update $s_{m-1}$  using  the stochastic approximation:
\begin{equation}\label{stoch_approx}
s_{m} =s_{m-1}+ a_{m-1} \left[S(\Vton,\Uton^{(m)})-s_{m-1}\right]
\end{equation}
   \item[\emph{M-Step:}] update of $\hthetam$  by $\hthetam=\underset{\theta \in \Theta}{\arg \max} \left(-\psi(\theta)+\langle s_{m},\nu(\theta)\rangle \right).$
\end{itemize}
\end{itemize}
\end{algo}
{Simulation under the smoothing distribution can be
  performed using a {\em naive forward} approach, which amounts to
  carry forward trajectories in the particle filter. We also 
  implemented a backward SMC smoother, with variance $O(n)$ instead
  of $O(n^2)$ for the naive smoother. However, in practice, the
  stochastic averaging of the SA step reduces the variance by 
  averaging over all the previous iterations using the step size
  $a_m$. }

Following \cite{Ditlevsen2014}, we can prove the convergence of the
SAEM-SMC algorithm, under standard  assumptions that are recalled in
the {Supplementary Material, Section
  \ref{App:assumptions}}. 

\begin{theorem}\label{thcvSAEM}
Assume that  (M1)-(M5),
(SAEM1)-(SAEM3), and (SMC1)-(SMC3) hold. Then, with probability 1,
$\lim_{m\rightarrow \infty}$ $ d( \hthetam, \mathcal{L}) = 0 $ where
$\mathcal{L}=\{ \theta \in \Theta, \partial_\theta \ell_\Delta
(\theta)=0\}$ is the set of stationary points of the log-likelihood
$\ell_\Delta(\theta) = \log \pDelta(\Vton;\theta)$.  

Moreover, under assumptions (LOC1)-(LOC3) given in \cite{Delyon1999}
on the regularity of the log-likelihood, the sequence $\hthetam$
converges with probability 1 to a (local) maximum of  the likelihood
$p_\Delta(\Vton;\theta)$.

\end{theorem}

The classical assumptions (M1)-(M5) are usually satisfied. Assumption (SAEM1) is easily satisfied by choosing properly the sequence $(a_m)$. Assumptions (SAEM2) and (SAEM3) depend on the regularity of the model. They are satisfied for the 3 approximate  models.

\subsection{Initializing the algorithm}\label{sec:initialization}
The SAEM algorithm requires initial values of $\theta$ to start. We
detail our strategy to find initial values for the two first
models. The SIE model is arbitrarily initialized with unknown
parameters fixed {at values of the correct order of magnitude}. 

For the HO model, we run the two-dimensional contrast based on complete observations of the two coordinates. As the $U$ coordinate is not observed, we replace it by the increments of $V$: $\tilde U_i = (V_{i+1}-V_i)/\Delta$. Then the two-dimensional criterion is minimized and initial values   $\hat D_0, \hat \gamma_0, \tilde \sigma_0$ are obtained. The value $\tilde \sigma_0$ is biased due to the approximation of $U_i$, as shown by \cite{Samson2012}. Therefore, we apply the bias correction suggested by \cite{Samson2012} and use $\hat \sigma_0= \sqrt{\frac32} \tilde \sigma_0$ as initial value. 

For the FHN model, the problem is more difficult because the unknown parameter $\varepsilon$ appears in the equation of the observed coordinate. We fix an arbitrary value for $\hat \varepsilon_0$. Then we replace the hidden coordinate $U_i$ by $\tilde U_i = V_{i}-V_i^3+s-\hat \varepsilon_0 \frac{V_{i+1}-V_i}{\Delta}$. Using $(V_i, \tilde U_i)$, we minimize the two-dimensional contrast to obtain initial values $\hat \gamma_0, {\hat \alpha _0}, \hat \varepsilon_0$.

\section{Simulation study}
\label{sec:simulation}

\subsection{Harmonic Oscillator}
Parameter values of the Harmonic Oscillator used in the simulations
are the same as those of \cite{Pokern2009, Samson2012}. The values
are: $D=4$, $\gamma=0.5$, $\sigma=0.5$. 
Trajectories are simulated with the exact distribution
eqs. \eqref{eq:HOdistribution1}--\eqref{eq:HOdistribution2}--\eqref{eq:HOdistribution3}
with time step
$\Delta=0.02$ and $n=1 000$ points.  Then $\theta$ is estimated on
each simulated trajectory.  
A hundred repetitions are used to evaluate the performance of the estimators.

The Particle filter aims at  filtering the hidden process $(U_t)$ from
the observed process $(V_t)$. We illustrate its performance on a
simulated trajectory, with $\theta$ fixed at its true value.
The SMC Particle filter algorithm is implemented with $K=100$ particles and the
conditional transition density as proposal.

The performance of the SAEM-SMC algorithm is illustrated on 100
simulated trajectories. The SAEM algorithm is implemented
with $m=80$ iterations and a sequence ($a_m$) equal to 1 during the 30
first iterations and equal to $a_m = 1/(m-30)^{0.9}$ for $m>
30$.  The SMC algorithm is implemented with $K(m)=100$ particles  at each
iteration of the SAEM algorithm. The
SAEM algorithm is initialized automatically by maximizing the log likelihood of the complete data, replacing the hidden $(U_{i\Delta})$ by the differences $((V_{(i+1)\Delta}-V_{i\Delta})/\Delta)$. 

Several  estimators are compared. The complete observation case is
illustrated with the new contrast estimator (numerical optimisation of
contrast (\ref{eq:hatphi})) and the Euler contrast from
\cite{Samson2012} (explicit estimators). The partial observation case
is illustrated with 
the SAEM estimator and the Euler contrast from
\cite{Samson2012}. Bayesian results from {the weak order 1.5 scheme} presented in \cite{Pokern2009} are also
recalled, even if they are obtained with a different sampling ($n=10
000$ and $\Delta=0.01$).  This estimator is known from  \cite{Pokern2009} to be biased. 
Results are given in Table \ref{tab:simuHO}. 

\begin{table}
\caption{\label{tab:simuHO} Harmonic Oscillator, mean and standard
  deviation (in parentheses) of estimators 
 calculated from 100
  trajectories with $\Delta=0.02$ and $n=1 \, 000$. Five estimation
  methods. Complete observations: new contrast estimator given in
  eq. \eqref{eq:hatphi} and Euler contrast from
  \cite{Samson2012}. Partial observations: SAEM, Euler contrast from
  \cite{Samson2012} and weak order 1.5 estimator from
  \cite{Pokern2009} obtained with 
  $n=10 \, 000$ and $\Delta=0.01$ (only the mean values for $D$ and
  $\gamma$ are given in their paper). } 

\begin{tabular}{cc|rr|rrr}
\hline
&& \multicolumn{4}{c}{ Observations}\\
   &  &  \multicolumn{2}{c|}{Complete}  & \multicolumn{3}{c}{Partial}\\
 \hline
   &True&New Contrast& Euler Contrast& SAEM & Euler Contrast & {weak order 1.5}\\
 \hline
  $D$& 4.0 & 3.712 (0.634) & 3.969 (0.540) &4.081 (0.503) & 3.969 (0.540) & 1.099 (--)\\
  $\gamma$ & 0.5& 0.701 (0.287) &  0.716 (0.273) &0.663 (0.273) & 0.754 (0.278) & 0.139 (--)\\
  $\sigma$ &0.5 & 0.496 (0.014)& 0.496 (0.011) & 0.509 (0.012) & 0.503 (0.011) & -- (--)\\
\hline
\end{tabular}
\end{table}

\medskip
The first four estimators give overall acceptable results, while the weak order 1.5 
estimator of \cite{Pokern2009} is seriously biased. 
The best results are obtained with the SAEM. It might seem surprising that
the SAEM  performs even better than the estimators based on complete
observations. This is due to the sensitivity of the numerical
optimisation of the contrast (\ref{eq:hatphi}) to the initial
conditions {for the iterative procedure}, that were set
to ${(\hat \gamma_0, \hat D_0, \hat \sigma_0)= }
(3,1,1)$. The stochasticity of the SAEM 
algorithm helps to avoid local optimization points, while the
numerical optimizer might get stuck in some local minimum. The
optimization of the Euler contrast is explicit for the HO
model, and there is thus no dependence on initial
conditions. It therefore outperforms the new contrast for $D$.   

Comparing the SAEM and the Euler contrast for the partial observation
case, they give results of the same order, even if slightly better for
the 
SAEM. However, the SAEM is much more time consuming. Note also that the
 SAEM algorithm provides confidence intervals easily, which is not
possible with the contrast estimators.

\subsection{FitzHugh-Nagumo model}

Parameter values of the FitzHugh-Nagumo model used in the simulations are : $\varepsilon=0.1$, $s=0$, $\gamma=1.5$, ${\alpha}=0.8$, $\sigma=0.3$.
Trajectories are simulated with time step $\delta=0.002$ and $n=1 000$ points are subsampled with observation time step $\Delta=10 \delta$.  Then $\theta$ is estimated on each simulated trajectory. 
A hundred repetitions are used to evaluate the performance of the estimators.

\begin{figure}[b!]
\centering
\includegraphics[width = 0.8\textwidth]{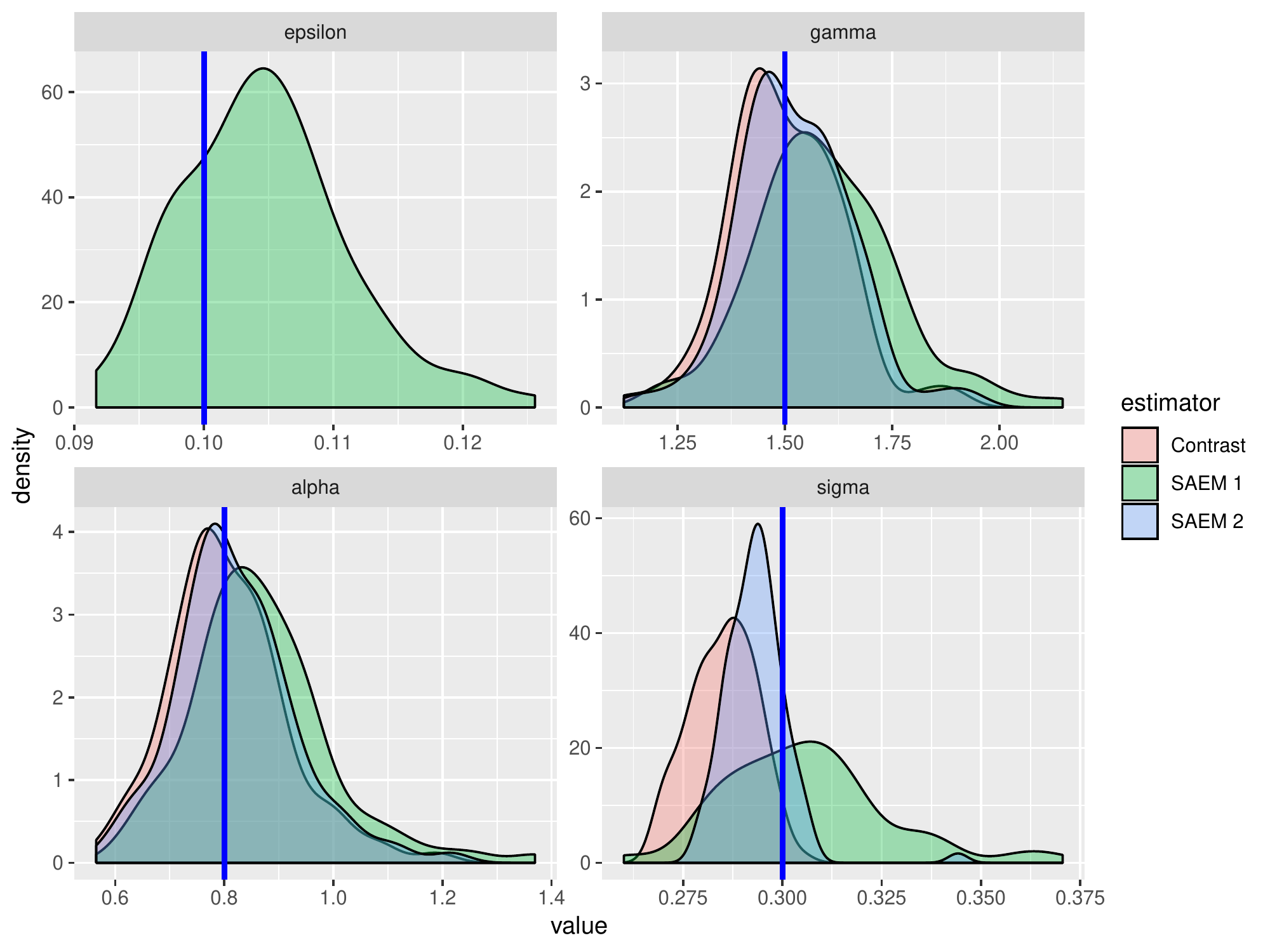}
\caption{\label{fig:FHNdensity} FHN estimation results for partial
  observations. Densities of estimated parameters over 100 repetitions
  for the new contrast method assuming $\varepsilon$ known (red), SAEM
  assuming $\varepsilon$ known (blue), SAEM estimating $\varepsilon$
  (green). The blue vertical lines are the true values. }
\end{figure}

Several  estimators are compared. First note that $\varepsilon$ is
difficult to estimate because it appears in the first
coordinate. Therefore, 
we first fix it at its true value. This allows to transform the system
into a Langevin equation with $dV_t=Z_tdt$, and to apply the Euler
contrast proposed by  \cite{Samson2012}. With $\varepsilon$ fixed, we
compare in the complete observation case the contrast estimator
(numerical optimisation of contrast (\ref{eq:hatphi})) and the Euler
contrast from \cite{Samson2012} (explicit estimators). We also include
the estimation of the full parameter vector by the new contrast given
in eqs. \eqref{eq:hatpsi} and \eqref{eq:hatphi}. In the partial
observation case we compare   
the SAEM estimator, the new contrast and the Euler contrast from \cite{Samson2012}.   We
also run the SAEM algorithm where $\varepsilon$ is not fixed but estimated.  
 
The SAEM algorithm is implemented
with $m=350$ iterations and a sequence ($a_m$) equal to 1 during the 250
first iterations and equal to $a_m = 1/(m-250)^{0.9}$ for $m>
250$.  The SMC algorithm is implemented with $K=100$ particles  at
each SAEM 
iteration. The
SAEM algorithm is initialized automatically by maximizing the log
likelihood of the complete data, replacing the hidden $(U_{i\Delta})$
by the differences $(V_{i\Delta}-V_{i\Delta}^3 s -
\varepsilon(V_{(i+1)\Delta}-V_{i\Delta}))/\Delta$, $\varepsilon$ being
initialized at ${\hat \varepsilon_0=}0.12$. Results are given in Table \ref{tab:simuFHN},
and densities of estimates in the partially observed case are
presented in Figure \ref{fig:FHNdensity}.

\begin{table}
\caption{\label{tab:simuFHN} FitzHugh-Nagumo model. 
Mean and standard deviation (in parentheses) of estimators
 calculated
from 100 trajectories with $\Delta=0.02$ and $n=1 \, 000$. Seven
estimation methods. Complete observations, $\varepsilon$ fixed: new
contrast estimator and Euler contrast from \cite{Samson2012}. Complete
observations, $\varepsilon$ estimated: new contrast estimator. Partial
observations, $\varepsilon$ fixed: SAEM, new contrast and Euler contrast from
\cite{Samson2012} $\varepsilon$ fixed.  Partial
observations, $\varepsilon$ estimated. SAEM. }
\begin{tabular}{lc|cccc}
\hline
   &  &  \multicolumn{3}{c}{Complete observations}  \\
 & &$\varepsilon$ fixed&$\varepsilon$ fixed&$\varepsilon$ estimated\\
  &True&New Contrast& Euler Contrast& New Contrast  \\
 \hline
  $\varepsilon$& 0.1& -- & --& 0.101 (0.0005) \\
  $\gamma$ &1.5&1.412 (0.221) &1.363 (0.201) &  1.516 (0.149)  \\
  ${\alpha}$ &0.8& 0.826 (0.146) &0.756 (0.131) & 0.822 (0.131)\\
  $\sigma$ &0.3&  0.303 (0.014) & 0.338 (0.024) & 0.299 (0.007)  \\
\hline
   &  &  
                                                       \multicolumn{3}{c}{Partial observations}\\
 & &$\varepsilon$ fixed &$\varepsilon$ fixed&$\varepsilon$ fixed& $\varepsilon$ estimated\\
  &True&SAEM & New Contrast & Euler Contrast  & SAEM\\
 \hline
  $\varepsilon$& 0.1& --& --& -- &0.105 (0.006) \\
  $\gamma$ &1.5&  1.523 (0.130) &1.512 (0.129)     &1.500 (0.130) &1.592 (0.165)  \\
  ${\alpha}$ &0.8& 0.822 (0.110) &0.815 (0.110) &0.807 (0.109) & 0.865 (0.129)\\
  $\sigma$ &0.3&  0.293 (0.008) & 0.300 (0.023)      &0.285 (0.008) &0.306 (0.021)  \\
\hline
\end{tabular}
\end{table}

The results are acceptable overall. In the complete observation case,
the new contrast gives better results than the Euler {contrast}. This is expected
because the new constrast has a higher order of convergence. For the
partial observation case, when $\varepsilon$ is fixed, the performance
of the SAEM and the
contrast are close. The Euler contrast gives better results with
partial observations than complete observations (except for
$\sigma$). This might be due to the sensitivity of the numerical
optimization used to minimize the criteria. Finally, the SAEM gives good
results when $\varepsilon$ is estimated, and this is the only method
that can estimate it.

\subsection{Synaptic-conductance model}
Parameter values of the SIE model used in the simulations are :
$G_L=50$, $V_L=-70$, $V_E=0$, $V_I=-80$, $I_{{inj}}=-60$, $\tau_E=0.5$,
$\tau_I=1$, $\bar g_E=17.8$, $\bar g_I=9.4$, $\sigma_E=0.1$,
$\sigma_I=0.1$. Initial conditions of the system are $V_0=-60$,
$G_{e,0}=10$, $G_{i,0}=1$.  

Trajectories are simulated with time step $\delta=0.002$ and $n=1 000$
points are subsampled with observation time step $\Delta=10 \delta$.
Then $\theta=(\tau_E, \tau_I, \bar g_E, \bar g_I, \sigma_E, \sigma_I)$
is estimated on each simulated trajectory.  
A hundred repetitions are used to evaluate the performance of the estimators.

The SAEM algorithm is implemented
with $m={80}$ iterations and a sequence ($a_m$) equal to 1 during the {30}
first iterations and equal to $a_m = 1/(m-{30})^{0.9}$ for $m>
{30}$.  The SMC algorithm is implemented with $K(m)=100$ particles  at each
iteration of the SAEM algorithm. The
SAEM algorithm is initialized  with {unknown parameters fixed at the
correct order of magnitude: time parameters are fixed to
1, unknown mean parameters are fixed to 10 and unknown standard deviation
parameters are fixed to 0.1.}

Results are given in Table \ref{tab:simuSIE}. { Parameters
$(\tau_E, \tau_I)$ are best estimated. Variances are larger for
estimates of the inhibitory parameters.} Inhibitory conductances
are generally more difficult to estimate, as also observed in
\cite{BergDitlevsen2013}, where analytic expressions for
approximations of 
the variance of the estimators of the conductances in a similar model
were derived from the Fisher Information matrix.  
This is because the dynamics of $V_t$ are close to the inhibitory
reversal potential $V_I$, whereas it is far from the excitatory
reversal potential $V_E$, and thus, the synaptic drive is higher for excitation.

\begin{table}
\caption{\label{tab:simuSIE} Synaptic conductance hypoelliptic model,
  estimation results obtained from 100 repeated trajectories with
  SAEM, from partial observations {(means and standard deviations over the 100 repeated trajectories)}.} 
\begin{tabular}{llrrrrrr}
\hline
&&\multicolumn{6}{c}{Parameters}\\
  && $\tau_E$ &  $\tau_I$ &$\bar g_E$&$\bar g_I$ & $\sigma_E$& $\sigma_I$\\
\hline
 true && 0.500& 1.000& 17.800& 9.400& 0.100& 0.100\\
\hline
mean &&    { 0.486} & { 0.990}& { 17.381} & { 8.414}&  { 0.076}&  { 0.098}  \\   
   {SD}&&{  0.031}& { 0.180}& { 0.110}& { 0.250}& { 0.003}& { 0.014}\\
 \hline
\end{tabular}
\end{table}

\section*{Acknowledgements}
\noindent   
Adeline Samson has been supported by the LabEx PERSYVAL-Lab
(ANR-11-LABX-0025-01). The work is part of the Dynamical Systems
Interdisciplinary Network, University of
Copenhagen. Villum Visiting Professor Programme funded a longer stay of
A. Samson at University of Copenhagen.


\newpage

\bigskip

\section{Supplementary material: Proofs of Propositions \ref{prop:consistency2} and \ref{prop:consistency1} and
  Theorems \ref{prop:normality} and \ref{prop:normality2} } \label{append_consistency}

To ease the notation, we assume that $p=1$ throughout this
Section. Furthermore, let $B_i(\theta) := B(X_i;\theta)$ and
$\Gamma_i(\sigma) := \Gamma (X_i;\sigma)$, and note that $\Gamma (\cdot)$ is a
scalar. Let $\nu(\cdot)$ denote the stationary density of model
\eqref{GeneralModel}. We write $\mathcal{G}_i$ for the filtration {generated} by $(X_t, t\leq t_i)$. 

\subsection{Technical lemmas}\label{sec:technicallemmas}
We first present the equivalent of Lemma 8-10 of \cite{Kessler1997}
that are essential for the proofs of
consistency. {The} 
equivalent of Lemma 7 is presented in Proposition \ref{prop:order_scheme}. 

\begin{lemma}\label{lemma8}
Let $f: \mathbb{R}^{p+1} \times \Theta \rightarrow \mathbb{R}$ be a function with derivatives of polynomial growth in $x$, uniformly in $\theta$. 
Assume $\Delta_n\rightarrow 0$ and $n\Delta_n \rightarrow \infty$. Then
$$ \nu_n(f):= \frac1n \sum_{i=1}^n f(X_{i}, \theta) \stackrel{P_{\theta_0}}{\rightarrow} \int f(x,\theta) \nu(dx) $$
uniformly in $\theta$.
\end{lemma}
The proof is the same as the proof of Lemma 8 in \cite{Kessler1997}.

\begin{lemma}\label{lemma9}
Let $f: \mathbb{R}^{p+1} \times \Theta \rightarrow \mathbb{R}$ be a
function with derivatives of polynomial growth in $x$, uniformly in
$\theta$.  
\begin{enumerate}
\item Assume $\Delta_n\rightarrow 0$ and $n \rightarrow \infty$. Then
\begin{eqnarray*}
Q_{1,n}(f):=\frac1{n\Delta_n^2} \sum_{i=0}^{n-1} f(X_{i}, \theta)(V_{i+1}-V_i - \Delta_n B_i(\theta_0)_1)^2 &\stackrel{P_{\theta_0}}{\rightarrow}& 0, \\
\end{eqnarray*}
uniformly in $\theta$.
\item Assume $\Delta_n\rightarrow 0$ and $n \rightarrow \infty$. Then
\begin{eqnarray*}
Q_{2,n}(f):=\frac1{n\Delta_n} \sum_{i=0}^{n-1} f(X_{i},
  \theta)(U_{i+1}-U_i - \Delta_n B_i(\theta_0)_2)^2
  &\stackrel{P_{\theta_0}}{\rightarrow}&  \int f(x,\theta) \Gamma^2(x;\sigma_0) \nu(dx) ,
\end{eqnarray*}
uniformly in $\theta$.
\end{enumerate}
\end{lemma}

\noindent \textbf{Proof of Lemma \ref{lemma9}}
To prove the first assertion (first coordinate), let 
$$\xi_{i+1}(\theta) = \frac1{n\Delta_n^2} f(X_{i}, \theta)(V_{i+1}-V_i - \Delta_n B_i(\theta_0)_1)^2$$
Due to Proposition  \ref{prop:order_scheme} and the ergodic theorem,
Lemma \ref{lemma8}, we have
\begin{eqnarray*}
\sum_{i=0}^{n-1}  \mathbb{E}_{\theta}(\xi_i(\theta)|\mathcal{G}_{i-1}) &=&
       \mathcal{O}  (\Delta_n) \rightarrow 0 \, \mbox{ for } \, \Delta_n
                                                                   \rightarrow 0\\
\sum_{i=0}^{n-1}  \mathbb{E}_{\theta}(\xi_i(\theta)^2|\mathcal{G}_{i-1}) &=&
      \frac1{n} \mathcal{O} \left (1\right ) \rightarrow 0 \, \mbox{ for } \, n
                                                                   \rightarrow \infty
\end{eqnarray*}
Hence, Lemma 9 from \cite{Genon-Catalot1993} proves the convergence for all $\theta$. 
Uniformity in $\theta$ follows as for Lemma \ref{lemma8}.  
The proof of the second assertion is the same. The scaling {(of
$n\Delta_n$)} is different { (from $n\Delta_n^2$)} because the variance of the scheme is of order $\Delta_n$ instead of order $\Delta_n^3$ (Proposition \ref{prop:order_scheme}). 
$\hfill \Box$

\begin{lemma}\label{lemma10}
Let $f: \mathbb{R}^{p+1} \times \Theta \rightarrow \mathbb{R}$ be a function with derivatives of polynomial growth in $x$, uniformly in $\theta$.

\begin{enumerate}
\item Assume $\Delta_n\rightarrow 0$ and $n\Delta_n \rightarrow \infty$.  Then
\begin{eqnarray*}
I_{1,f}:=\frac1{n\Delta_n^2} \sum_{i=0}^{n-1} f(X_{i}, \theta)(V_{i+1}-V_i - \Delta_n B_i(\theta_0)_1) &\stackrel{P_{\theta_0}}{\rightarrow}& 0, \\
\end{eqnarray*}
uniformly in $\theta$
\item Assume $\Delta_n\rightarrow 0$ and $n\Delta_n \rightarrow \infty$.  Then 
\begin{eqnarray*}
I_{2,f}:=\frac1{n\Delta_n} \sum_{i=0}^{n-1} f(X_{i},
  \theta)(U_{i+1}-U_i - \Delta_n B_i(\theta_0)_2) & \stackrel{P_{\theta_0}}{\rightarrow}&0,
\end{eqnarray*}
uniformly in $\theta$
\item Assume $\Delta_n\rightarrow 0$ and $n \rightarrow \infty$.  Then 
\begin{eqnarray*}
I_{3,f}:=\frac1{n} \sum_{i=0}^{n-1} f(X_{i},
  \theta)(U_{i+1}-U_i - \Delta_n B_i(\theta_0)_2) & \stackrel{P_{\theta_0}}{\rightarrow}&0,
\end{eqnarray*}
uniformly in $\theta$
\end{enumerate}
\end{lemma}

\noindent \textbf{Proof of Lemma \ref{lemma10}}
To prove the first assertion (first coordinate), let 
$$\xi_{i+1}(\theta) = \frac1{n\Delta_n^2} f(X_{i}, \theta)(V_{i+1}-V_i - \Delta_n B_i(\theta_0)_1)$$
Due to Proposition  \ref{prop:order_scheme} and intermediate calculations (not shown), we have
\begin{eqnarray*}
\sum_{i=0}^{n-1}  \mathbb{E}_{\theta_0}(\xi_i(\theta)|\mathcal{G}_{i-1}) &=&  
\mathcal{O}(\Delta_n)  \rightarrow 0 \, \mbox{ for } \, \Delta_n
                                                                   \rightarrow 0\\
\sum_{i=0}^{n-1}  \mathbb{E}_{\theta_0}(\xi_i(\theta)^2|\mathcal{G}_{i-1}) &=& \frac{1}{n \Delta_n}\mathcal{O} \left (1\right )  \rightarrow 0 \, \mbox{ for } \, n\Delta_n
                                                                   \rightarrow \infty
\end{eqnarray*}
Hence, Lemma 9 from \cite{Genon-Catalot1993} proves the convergence for all $\theta$. The proof of uniformity in $\theta$ is the same as for Lemma 10 of \cite{Kessler1997}. 

The proofs of the second and third assertions are the same, only the
scalings are different due to Proposition \ref{prop:order_scheme}. 
$\hfill \Box$

Next we present some Lemmas which are needed to prove asymptotic normality.

  \begin{lemma}\label{lemma:Th45Gloter}
  \begin{enumerate}
\item   Assume that $n\Delta_n^2\rightarrow 0$. Then
  $$\frac1{\sqrt{n\Delta_n}} \sum_{i=0}^{n-1}f(X_i)(U_{i+1}-U_i - \Delta_n B_i(\theta_0)_2) \stackrel
                                     {\mathcal{D}}{\rightarrow} \mathcal{N}(0, \nu(\Gamma^2 f^2(\cdot)))$$
  
  \item Assume that $n\Delta_n^2\rightarrow 0$. Then 
  $$\frac1{\sqrt{n}\Delta_n} \sum_{i=0}^{n-1} f(X_i)(U_{i+1}-U_i)^2 -
  \frac1{\sqrt{n}} \sum_{i=0}^{n-1} f(X_i)\Gamma^2_i(\sigma_0)
  \stackrel {\mathcal{D}}{\rightarrow} \mathcal{N}(0,
  2\nu((\Gamma (\cdot))^4 f^2(\cdot)))$$ 

  \end{enumerate}
  \end{lemma}
  
\paragraph{Proof of Lemma \ref{lemma:Th45Gloter}.} 
Recall that $U_{i+1}-U_i  - \Delta_n
B_i(\theta_0)_2 = \sqrt{\Delta_n} \tilde\xi_i^U \Gamma_i(\sigma_0) +
\epsilon_i^U$, where $\sqrt{\Delta_n} \tilde\xi_i^U=\eta_i + \partial_u
A \xi_i$ and $\epsilon_i^U$
is the difference between the true process and the scheme. Thus, $\mathbb{E}(\tilde\xi_i^U)=0$,
$Var(\tilde\xi_i^U)=1+\mathcal{O}(\Delta_n)$, $Cov(\xi_i^U,
\xi_{i+1}^U)=0$, and from Proposition \ref{prop:order_scheme}, {it} follows that 
$\mathbb{E}(\epsilon_i^U) = \mathcal{O} (\Delta_n^3)$ and $Var(\epsilon_i^U) = \mathcal{O} (
\Delta_n^2)$.  To prove assertion a), rewrite 
\begin{eqnarray*}
\frac1{\sqrt{n\Delta_n}} \sum_{i=0}^{n-1}\! f(X_i)(U_{i+1}\! -\! U_i \! -
  \! \Delta_n
  B_i(\theta_0)_2)\! &= &\! \! \frac{\sqrt{\Delta_n}}{\sqrt{n\Delta_n}} \sum_{i=0}^{n-1}
                       \tilde\xi_i^U  \Gamma_i (\theta)
                       f(X_i) + \! \frac1{\sqrt{n\Delta_n}} \sum_{i=0}^{n-1} \epsilon_i^U f(X_i)\\
& =& \! T_1+T_2
\end{eqnarray*}
Since $\mathbb{E}(\tilde\xi_i^U
\Gamma_i(\theta) f(X_i)  | \mathcal{G}_i)=0$ and 
$\mathbb{E}((\tilde\xi_i^U \Gamma_i(\theta) f(X_i))^2  |
\mathcal{G}_i)=(\Gamma_i(\theta))^2 f(X_i)^2(1+\mathcal{O}(\Delta_n))$, then $\frac1n
\sum_{i=0}^{n-1}\mathbb{E}\left( (\tilde\xi_i^U \Gamma_i(\theta)
  f(X_i))^2|\mathcal{G}_i\right) \rightarrow \nu(\Gamma^2(\cdot,
\theta) f(\cdot)^2) $. Since $\mathbb{E}((\tilde\xi_i^U)^4
(\Gamma_i( \theta))^4 f(X_i)^4  | \mathcal{G}_i)$ is bounded it follows that
$\frac1{n^2} \sum_{i=0}^{n-1} \mathbb{E}((\tilde\xi_i^U)^4
(\Gamma_i(\theta))^4 f(X_i)^4  |
\mathcal{G}_i)\rightarrow0$. Using theorem 3.2 in \cite{Hall1980},
these two conditions are sufficient to imply 
$$T_1 = \frac1{\sqrt{n}}\sum_{i=0}^{n-1}  \tilde\xi_i^U  \Gamma_i(\sigma)
f(X_i)\stackrel {\mathcal{D}}{\rightarrow} \mathcal{N}(0, \nu(f^2\Gamma^2)). $$ 
Then we study $T_2$. We have $\frac1{\sqrt{n\Delta_n}}
\sum_{i=0}^{n-1} \mathbb{E}(\epsilon_i^U|\mathcal{G}_i)= \sqrt{n}\mathcal{O}(\sqrt{\Delta_n^5})$
and $\frac1{n\Delta_n} \sum_{i=0}^{n-1}
\mathbb{E}((\epsilon_i^U)^2|\mathcal{G}_i)=\mathcal{O}( \Delta_n)$. The condition
$n\Delta^2_n\rightarrow0$ implies $n\Delta^5_n\rightarrow0$ and $T_2\rightarrow 0$. This gives
the proof of 1.  

\medskip
To prove assertion 2, 
rewrite
\begin{align*}
\frac1{\sqrt{n}\Delta_n} \sum_{i=0}^{n-1}& f(X_i)\left((U_{i+1}-U_i)^2
   - \Delta_n \Gamma^2_i(\sigma_0)\right)=\frac{1}{\sqrt{n} }
   \sum_{i=0}^{n-1}  \Gamma^2_i(\sigma_0)
   ((\tilde\xi_i^U)^2-1)f(X_i)\\ 
 +& \frac{2}{\sqrt{n \Delta_n}} \sum_{i=0}^{n-1}   (\epsilon_i^U +
   \Delta_n B_i(\theta_0)_2) \Gamma_i(\sigma_0) \tilde\xi_i^Uf(X_i) +
   \frac{1}{\sqrt{n} \Delta_n} \sum_{i=0}^{n-1} (\epsilon_i^U +
   \Delta_n  B_i(\theta_0)_i)^2f(X_i)\\ 
=& T_1+T_2+T_3
\end{align*}
Note that $\mathbb{E}((\tilde\xi_i^U)^2-1|\mathcal{G}_i)=\mathcal{O}(\Delta_n)$
and
$\mathbb{E}(((\tilde\xi_i^U)^2-1)^2|\mathcal{G}_i)=2+\mathcal{O}(\Delta_n)$. Thus, 

\noindent $\displaystyle{\frac1n \sum_{i=0}^{n-1}}\mathbb{E}\left( \left (
  \Gamma^2_i(\sigma_0)
  ((\tilde\xi_i^U)^2-1)f(X_i)\right )^2|\mathcal{G}_i\right)\rightarrow
2\nu(\Gamma^4(\cdot, \theta) f(\cdot)^2) $.  Since

\noindent $\mathbb{E}\left ( \left (((\tilde\xi_i^U)^2
    -1)(\Gamma^2_i(\theta)) f(X_i)\right )^4
| \mathcal{G}_i\right )$ is bounded it follows that 

\noindent $\frac1{n^2} \sum_{i=0}^{n-1}
\mathbb{E}\left (\left (((\tilde\xi_i^U)^2 -1) \Gamma^2_i(\theta) f(X_i)
\right )^4
| \mathcal{G}_i \right )\rightarrow0$. Using theorem 3.2 in \cite{Hall1980},
these two conditions are sufficient to imply 
$$T_1 = \frac{1}{\sqrt{n}} \sum_{i=0}^{n-1}  \Gamma^2_i(\sigma_0) ((\tilde\xi_i^U)^2-1)f(X_i)\stackrel {\mathcal{D}}{\rightarrow}  \mathcal{N}(0, 2\nu(f^2\Gamma^4)). $$
We have
$\frac1{n\Delta_n} \sum_{i=0}^{n-1} \mathbb{E}((\epsilon_i^U + \Delta_n
B_i(\theta_0)_2)^2 \Gamma^2_i(\sigma_0) (\tilde\xi_i^U)^2f^2(X_i)|\mathcal{G}_i)
= \mathcal{O} (\Delta_n^2)$ goes to $0$ when $\Delta_n \rightarrow 0$
since $\mathbb{E}((\tilde\xi_i^U)^2(\epsilon_i^U + \Delta_n 
B_i(\theta_0)_2)^2 |\mathcal{G}_i)=\mathcal{O}( \Delta_n^2)$, which implies $T_2\rightarrow 0$. 
Furthermore, the condition
$n\Delta_n^2\rightarrow 0$ and $\mathbb{E}((\epsilon_i^U+\Delta_n
B_i(\theta_0)_2)^2|\mathcal{G}_i) =\mathcal{O} (\Delta_n^2)$ imply $\mathbb{E}(T_3)\rightarrow 0$. We also have $\mathbb{E}((\epsilon_i^U+\Delta_n
B_i(\theta_0)_2)^4|\mathcal{G}_i) =\mathcal{O} (\Delta_n^3)$. We can conclude that $T_3\rightarrow 0$.  
This proves Lemma \ref{lemma:Th45Gloter}. $\hfill \Box$

\begin{lemma}\label{lemma:Th45GloterV}
  \begin{enumerate}
\item   Assume that $n\Delta_n^2\rightarrow 0$. Then
  $$\frac1{\sqrt{n\Delta_n^3}} \sum_{i=0}^{n-1}f(X_i)(V_{i+1}-V_i - \Delta_n B_i(\theta_0)_1) \stackrel
                                     {\mathcal{D}}{\rightarrow}
                                     \mathcal{N}(0, \frac13
                                     \nu((\partial_u a)^2\Gamma^2 f^2(\cdot)))$$
  
  \item Assume that $n\Delta_n^2\rightarrow 0$. Then 
  $$\frac1{\sqrt{n}\Delta_n^3} \sum_{i=0}^{n-1}
  f(X_i)(V_{i+1}-V_i-\Delta_n B_i(\theta)_1)^2 -
  \frac1{\sqrt{n}} \sum_{i=0}^{n-1} f(X_i)\frac13
  \Gamma^2_i(\sigma_0) (\partial_u a)^2$$ 
  $$\stackrel {\mathcal{D}}{\rightarrow} \mathcal{N}(0,
  \frac29\nu(\Gamma^4 (\partial_u a)^4 f^2(\cdot)))$$ 

  \end{enumerate}
  \end{lemma}
  
\paragraph{Proof of Lemma \ref{lemma:Th45GloterV}.} 
Recall that $V_{i+1}-V_i  - \Delta_n
B_i(\theta_0)_1 = \sqrt{\Delta_n^3} \tilde\xi_i^V \Gamma_i +
\epsilon_i^V$, where $\sqrt{\Delta_n^3} \tilde\xi_i^V=\partial_u a \xi_i$ and $\epsilon_i^V$
is the difference between the true process and the scheme. Thus, $\mathbb{E}(\tilde\xi_i^V)=0$,
$Var(\tilde\xi_i^V)=\frac13 (\partial_u a)^2$, $Cov(\xi_i^V,
\xi_{i+1}^V)=0$, and from Proposition \ref{prop:order_scheme}, {it} follows that 
$\mathbb{E}(\epsilon_i^V) = \mathcal{O} (\Delta_n^3)$ and $Var(\epsilon_i^V) = \mathcal{O} (
\Delta_n^4)$.  To prove assertion a), rewrite 
\begin{eqnarray*}
\frac1{\sqrt{n\Delta_n^3}} \sum_{i=0}^{n-1}\! f(X_i)(V_{i+1}\! -\! V_i \! -
  \! \Delta_n
  B_i(\theta_0)_1)\! &= &\! \! \frac{\sqrt{\Delta_n^3}}{\sqrt{n\Delta_n^3}} \sum_{i=0}^{n-1}
                       \tilde\xi_i^V  \Gamma_i (\theta)
                       f(X_i) + \! \frac1{\sqrt{n\Delta_n^3}} \sum_{i=0}^{n-1} \epsilon_i^V f(X_i)\\
& =& \! T_1+T_2
\end{eqnarray*}
Note that $\mathbb{E}(\tilde\xi_i^V
\Gamma_i(\theta) f(X_i)  | \mathcal{G}_i)=0$ and 
$\mathbb{E}((\tilde\xi_i^V\Gamma_i(\theta)f(X_i))^2  |
\mathcal{G}_i)=\frac13(\partial_u a\Gamma_i(\theta) f(X_i))^2$. Thus, 
$\frac1n
\sum_{i=0}^{n-1}\mathbb{E}\left( (\tilde\xi_i^V \Gamma_i(\theta)
  f(X_i))^2|\mathcal{G}_i\right) \rightarrow \frac13 \nu((\partial_u a)^2\Gamma^2(\cdot,
\theta) f(\cdot)^2) $. Since $\mathbb{E}((\tilde\xi_i^V \Gamma_i(
\theta) f(X_i))^4  | \mathcal{G}_i)$ is bounded it follows that 
$\frac1{n^2} \sum_{i=0}^{n-1} \mathbb{E}((\tilde\xi_i^V
\Gamma_i(\theta) f(X_i))^4  |
\mathcal{G}_i)\rightarrow0$. Using theorem 3.2 in \cite{Hall1980},
these two conditions are sufficient to imply 
$$T_1 = \frac1{\sqrt{n}}\sum_{i=0}^{n-1}  \tilde\xi_i^V  \Gamma_i(\sigma)
f(X_i)\stackrel {\mathcal{D}}{\rightarrow} \mathcal{N}(0, \frac13 \nu(f^2(\partial_u a)^2\Gamma^2)). $$ 
To study $T_2$, note that $\frac1{\sqrt{n\Delta_n^3}}
\sum_{i=0}^{n-1} \mathbb{E}(\epsilon_i^V|\mathcal{G}_i)= \sqrt{n}\mathcal{O}(\sqrt{\Delta_n^3})$
and $\frac1{n\Delta_n^3} \sum_{i=0}^{n-1}
\mathbb{E}((\epsilon_i^V)^2|\mathcal{G}_i)=\mathcal{O}( \Delta_n)$. The condition
$n\Delta^2_n\rightarrow0$ implies $n\Delta^3_n\rightarrow0$  and $T_2\rightarrow 0$. This gives
the proof of a).  

\medskip
To prove assertion b), 
rewrite
\begin{align*}
\frac1{\sqrt{n}\Delta_n^3} \sum_{i=0}^{n-1} &f(X_i)\left((V_{i+1}-V_i-\Delta_n B_i(\theta_0)_1)^2
   - \Delta_n^3 \frac13 (\partial_u a)^2 \Gamma^2_i(\sigma_0)\right)\\
=&\frac{1}{\sqrt{n} }
   \sum_{i=0}^{n-1}  \Gamma^2_i(\sigma_0)
   ((\tilde\xi_i^V)^2-\frac13(\partial_u a)^2)f(X_i)\\ 
& + \frac{2}{\sqrt{n \Delta_n^3}} \sum_{i=0}^{n-1} \epsilon_i^V
   \Gamma_i(\sigma_0) \tilde\xi_i^Vf(X_i) + 
   \frac{1}{\sqrt{n} \Delta_n^3} \sum_{i=0}^{n-1} (\epsilon_i^V)^2f(X_i)\\ 
=& T_1+T_2+T_3
\end{align*}
Note that $\mathbb{E}((\tilde\xi_i^V)^2-\frac13 (\partial_u a)^2|\mathcal{G}_i)=0$
and
$\mathbb{E}(((\tilde\xi_i^V)^2-\frac13 (\partial_u a)^2)^2|\mathcal{G}_i)=\frac29 (\partial_u a)^4$. Thus, 
$\frac1n \sum_{i=0}^{n-1}\mathbb{E}\left( (
  \Gamma^2_i(\sigma_0)
  ((\tilde\xi_i^V)^2-\frac13 (\partial_u a)^2)f(X_i))^2|\mathcal{G}_i\right)\rightarrow
\frac29 \nu(\Gamma^4(\cdot, \theta) (\partial_u a)^4 f(\cdot)^2) $.
Moreover, since
$\mathbb{E}(((\tilde\xi_i^V)^2 -\frac13 (\partial_u a)^2))^4(\Gamma^2_i(\theta))^4 f(X_i)^4 
| \mathcal{G}_i))$ is bounded, it follows that 

\noindent $\frac1{n^2} \sum_{i=0}^{n-1}
\mathbb{E}(((\tilde\xi_i^V)^2 -\frac13 (\partial_u a)^2))^4 (\Gamma^2_i(\theta))^4 f(X_i)^4
| \mathcal{G}_i))\rightarrow0$. Using theorem 3.2 in \cite{Hall1980},
these two conditions are sufficient to imply 
$$T_1 = \frac{1}{\sqrt{n} } \sum_{i=0}^{n-1} \Gamma^2_i(\sigma_0) ((\tilde\xi_i^V)^2-\frac13 (\partial_u a)^2))f(X_i)\stackrel {\mathcal{D}}{\rightarrow}  \mathcal{N}(0, \frac29 \nu(\Gamma^4(\cdot, \theta) (\partial_u a)^4 f(\cdot)^2)). $$
We have
$\frac1{n\Delta_n^3} \sum_{i=0}^{n-1} \mathbb{E}((\epsilon_i^V)^2
\Gamma^2_i(\sigma_0) (\tilde\xi_i^V)^2f^2(X_i)|\mathcal{G}_i) 
= \mathcal{O} (\Delta_n)$ goes to $0$ when $\Delta_n \rightarrow 0$
since $\mathbb{E}((\tilde\xi_i^V)^2(\epsilon_i^V)^2 |\mathcal{G}_i)=\mathcal{O}(
\Delta_n^4)$, which implies $T_2\rightarrow 0$.  
Furthermore, the condition
$n\Delta_n^2\rightarrow 0$ and $\mathbb{E}((\epsilon_i^V)^2|\mathcal{G}_i)
=\mathcal{O} (\Delta_n^4)$ imply $T_3\rightarrow 0$.   
This proves Lemma \ref{lemma:Th45GloterV}. $\hfill \Box$

\subsection{Proof of consistency of $\hat \sigma_n^2$, Proposition \ref{prop:consistency1}}

The estimator  $\hat \sigma_n^2$ is  defined as the minimal argument
  of (\ref{eq:hatphi}) which for $p=1$ reduces to  
\begin{equation}\label{eq:contrast2}
\ell_n(\beta, \sigma)= \sum_{i=0}^{n-1}\frac{ \left(U_{i+1}-U_i -
    \Delta_n B_i(\beta)_2\right)^2}{ \Delta_n \Gamma^2_i(\sigma)} +
\sum_{i=0}^{n-1}\log (\Gamma^2_i(\sigma)).
\end{equation} 
We follow
\cite{Kessler1997} and the aim is to prove the following lemma 
\begin{lemma}\label{lemma:contrastsigma} Assume $\Delta_n\rightarrow 0$ and $n \Delta_n \rightarrow \infty$.  Then 
\begin{equation}\label{eq:contrastsigma}
 \frac1n \ell_n(\beta, \sigma) \stackrel{P_{\beta_0}}{\rightarrow}
 \int \left( \frac{\Gamma^2(x; \sigma_0)}{\Gamma^2(x; \sigma)}
   +\log \Gamma^2(x; \sigma)\right) \nu(dx)=: F(\sigma,
 \sigma_0) 
 \end{equation}
uniformly in $\theta$.  \end{lemma}
Then, using  Lemma \ref{lemma:contrastsigma}, we can prove that there exists a subsequence $n_k$ such that $(\hat \varphi_{n_k}, \hat \sigma_{n_k})$ converges to a limit $(\varphi_\infty, \sigma^2_\infty)$. Hence, by continuity of $\sigma \rightarrow F(\sigma, \sigma_0)$, we have 
$$ \frac1{n_k} \ell_{n_k}(\beta, \sigma)  \stackrel{P_{\theta_0}}{\rightarrow} F(\sigma_\infty, \sigma_0). $$
By definition of $(\hat \varphi_{n_k}, \hat \sigma_{n_k})$, $F(\sigma_\infty, \sigma_0)\leq F(\sigma_0, \sigma_0)$.

On the other hand,   for all $y>0, y_0>0, (y_0/y)+\log y\geq 1+\log
y_0$. Thus, $F(\sigma_\infty, \sigma_0)= F(\sigma_0,
\sigma_0)$, and by identifiability assumption
$\sigma_\infty^2=\sigma_0^2$. Hence, there exists a subsequence of
$\hat\sigma_n^2$ that converges to $\sigma^2_0$. That proves the
consistency of $\hat\sigma_n^2$. It remains to prove  Lemma
\ref{lemma:contrastsigma}.

\paragraph{Proof of Lemma \ref{lemma:contrastsigma}} We have $ \frac1n \ell_n(\beta, \sigma) =T_1+T_2+T_3+T_4 $
with
\begin{eqnarray*}
T_1&=&\frac1n \sum_{i=0}^{n-1} \frac{(U_{i+1}-U_i-  \Delta_n B_i(\beta_0)_2)^2}{\Delta_n  \Gamma^2_i(\sigma)}\\
T_2&=&\frac{2}n \sum_{i=0}^{n-1} \frac{(U_{i+1}-U_i-\Delta_n B_i(\beta_0)_2)( B_i(\beta_0)_2-B_i(\beta)_2)}{\Gamma^2_i(\sigma)}\\
T_3&=& \frac{\Delta_n}n \sum_{i=0}^{n-1}\frac{(
       B_i(\beta_0)_2-
       B_i(\beta)_2)^2}{\Gamma^2_i (\sigma)}\\
T_4&=& \frac1n \sum_{i=0}^{n-1} \log \Gamma^2_i (\sigma) 
\end{eqnarray*}
We start with $T_1$.  Lemma \ref{lemma9} implies 
$$\frac1{n\Delta_n}  \sum_{i=1}^{n-1} (U_{i+1}-U_i-\Delta_n B_i(\beta_0)_2)^2\stackrel{P_{\theta_0}}{\rightarrow}   \int \Gamma^2(x; \sigma_0)\nu(dx)$$
and thus, $T_1\stackrel{P_{\theta_0}}{\rightarrow} \int
\frac{\Gamma^2(x; \sigma_0)}{\Gamma^2(x; \sigma)}\nu(dx)$,
uniformly in $\theta$.  
Using Lemma \ref{lemma10}, we obtain that
$T_2\stackrel{P_{\theta_0}}{\rightarrow} 0$, uniformly in
$\theta$. From Lemma \ref{lemma8} follows
$T_3\stackrel{P_{\theta_0}}{\rightarrow} 0$ and
$T_4\stackrel{P_{\theta_0}}{\rightarrow} \int \log \Gamma^2(x; \sigma) \nu(dx)$, uniformly in
$\theta$.  
Finally, we obtain (\ref{eq:contrastsigma}). $\hfill \Box$

\subsection{Proof of consistency of $\hat \varphi_n$, Proposition \ref{prop:consistency1}}

The estimator  $\hat \varphi_n$ is  defined as the minimal argument of \eqref{eq:contrast2}. 
Consistency of $\hat \varphi_n$ is deduced from the following lemma.
\begin{lemma}\label{lemma:seconddrift}
Assume $\Delta_n\rightarrow 0$ and $n\Delta_n \rightarrow \infty$.  Then 
$$\frac1{n\Delta_n} \ell_n(\beta, \sigma)- \frac1{n\Delta_n} \ell_n(\beta_0, \sigma)\stackrel{P_{\theta_0}}{\rightarrow}  \int\frac{ \left(A(x;\varphi)-A(x;\varphi_0)\right)^2}{\Gamma^2(x; \sigma)} \nu(dx) $$
uniformly in $\theta$. 
\end{lemma}

Using Lemma \ref{lemma:seconddrift}, there exists a subsequence $\hat
\varphi_{n_k}$ that tends to $\varphi_\infty$. Hence, 
$$\frac1{n_k\Delta_{n_k}} \ell_{n_k}(\hat \beta_{n_k}, \sigma)- \frac1{n_k\Delta_{n_k}} \ell_{n_k}(\beta_0, \sigma) \stackrel{P_{\theta_0}}{\rightarrow} \int \frac{ \left(A(x;\varphi_{\infty})-A(x;\varphi_0)\right)^2}{\Gamma^2(x; \sigma)}  \nu(dx)$$
The consistency follows by identifiability of $A(x;\varphi)$.  It remains to prove Lemma \ref{lemma:seconddrift}.

\noindent\textbf{Proof of Lemma \ref{lemma:seconddrift}}. 
 We have $\frac1{n\Delta_n} \ell_n(\beta, \sigma)- \frac1{n\Delta_n} \ell_n(\beta_0, \sigma)=T_1+T_2$ with
\begin{eqnarray*}
T_1&=&  \frac{2}{n\Delta_n } \sum_{i=0}^{n-1} \frac{(U_{i+1}-U_i -\Delta_n B_i(\beta_0)_2)}{\Gamma^2_i(\sigma)}(B_i(\beta_0)_2-B_i(\beta)_2)  \\
T_2&=& \frac1{n } \sum_{i=0}^{n-1} \frac{( B_i(\beta_0)_2- B_i(\beta)_2)^2}{\Gamma^2_i(\sigma)} \\
\end{eqnarray*}
Lemma \ref{lemma10} implies $T_1\stackrel{P_{\theta_0}}{\rightarrow} 0$, uniformly in
$\theta$.  Recall that  $B_i(\beta_0)_2-B_i(\beta)_2 = A(X_i;\varphi_0)-A(X_i;
\varphi) + \mathcal{O}(\Delta_n)$. Combined with Lemma \ref{lemma8} we obtain $
T_2\stackrel{P_{\theta_0}}{\rightarrow} \int\frac{
  \left(A(x;\varphi)-A(x;\varphi_0)\right)^2}{\Gamma^2(x;
  \sigma)} \nu(dx)$, uniformly in $\theta$.  Note that the parameter of the first coordinate $\psi$ is not involved in the limit. The result applies for any $\psi$. This gives the Lemma.  
$\hfill \Box$

\subsection{Proof of consistency of $\hat \psi_n$, Proposition \ref{prop:consistency2}}

Assume that the drift function
$a$ can be split into two functions of $v$ and $u$: $a(x;\psi)= a_v(v,\psi_v) +
\psi_u a_u(u)$.  
Estimator $\hat \psi_n= (\hat{\psi_v}_n, \hat{\psi_u}_n)$   is
defined as the minimal argument of (\ref{eq:hatpsi}) which for $p=1$ reduces to 
\begin{eqnarray}
\label{eq:contrast3}
\ell_n(\psi, \sigma) &=& \frac3{\Delta_n^3}
                           \sum_{i=0}^{n-1}\frac{(V_{i+1}-V_i-\Delta_n
                           B_i(\beta )_1)^2}{  \psi_u^2 \,
                           \Gamma^2_i(\sigma) (a'_u(U_i))^2} +
                           n\log(\psi_u^2).
\end{eqnarray} 
Consistency of $\hat \psi_n$ is deduced from the following lemma.
\begin{lemma}\label{lemma:firstdrift}
Assume $\Delta_n\rightarrow 0$ and $n\Delta_n\rightarrow \infty$. Then 
$$\frac{\Delta_n}{n} \ell_n(\psi, \sigma)- \frac{\Delta_n}{n}
\ell_n(\psi_0, \sigma)\stackrel{P_{\theta_0}}{\rightarrow}
\int\frac{ \left(a(x;\psi)-a(x;\psi_0)\right)^2}{\psi_u^2\,
  \Gamma^2(x; \sigma)(a'_u(u))^2} \nu(dx) $$ 
uniformly in $\theta$. 
\end{lemma}
\noindent\textbf{Proof of Lemma \ref{lemma:firstdrift}}. 
 We have $\frac{\Delta_n}{n} \ell_n(\psi, \sigma)- \frac{\Delta_n}{n} \ell_n(\psi_0, \sigma)=T_1+T_2+T_3+T_4$ with
\begin{eqnarray*}
T_1&=& \frac{3\Delta_n}{n\Delta_n^3}\sum_{i=0}^{n-1}\frac{(V_{i+1}-V_i
       -\Delta_n B_i(\beta_0)_1)^2}{\Gamma^2_i
       (\sigma) (a'_u(U_i))^2}\left(\frac1{\psi_u^2\, }-\frac1{\psi_{u,0}^2\,
       }\right)\\ 
T_2&=& \frac{6\Delta_n^2}{n\Delta_n^3 } \sum_{i=0}^{n-1} \frac{(V_{i+1}-V_i -\Delta_n B_i(\beta_0)_1)}{\Gamma^2_i(\sigma) (a'_u(U_i))^2}\frac{(B_i(\beta_0)_1-B_i(\beta)_1)}{\psi_u^2}  \\
T_3&=& \frac{3\Delta_n^3}{n\Delta_n^3 } \sum_{i=0}^{n-1} \frac{(
       B_i(\beta_0 )_1-B_i(\beta)_1)^2}{\psi_u^2\,
       \Gamma^2_i(\sigma) (a'_u(U_i))^2} \\ 
T_4&=& \Delta_n \log(\psi_u^2/\psi_{u,0}^2)
\end{eqnarray*}
Lemma
\ref{lemma9} implies $T_1\stackrel{P_{\theta_0}}{\rightarrow}
0$ and Lemma \ref{lemma10} implies
$T_2\stackrel{P_{\theta_0}}{\rightarrow} 0$, uniformly in $\theta$.
From Lemma
\ref{lemma8} combined with $B_{i}(\beta_0)_1- B_{i}(\beta)_1 =
a(X_i;\psi_0)-a(X_i;\psi) + \mathcal{O}(\Delta_n)$ follows that  $
T_3\stackrel{P_{\theta_0}}{\rightarrow}  3  \int\frac{
  \left(a(x;\psi)-a(x;\psi_0)\right)^2}{\psi_u^2\,
  \Gamma^2(x;\sigma) (a'_u(u))^2} \nu(dx)$, uniformly in $\theta$. Finally
$T_4\stackrel{P_{\theta_0}}{\rightarrow} 0$, uniformly in
$\theta$. Note that the parameter of the second coordinate $\varphi$ is not involved in the limit. The result applies for any $\varphi$.  This gives the Lemma.  
$\hfill \Box$

\subsection{Proof of the asymptotic normality of $(\hat \varphi_n, \hat \sigma_n^2)$ (Theorem \ref{prop:normality})}

\paragraph{Proof of Theorem \ref{prop:normality}.} 
 The proof of the asymptotic normality is standard, see for instance
 \cite{Genon-Catalot1993, Kessler1997}. Denote $\theta=(\psi, \varphi,
 \sigma)$ and $\hat \theta_n=(\psi_0, \hat \varphi_n, \hat \sigma_n)$.  Let
 $\mathcal{L}_n(\theta) =  \ell_n(\beta, \sigma)$ from
 \eqref{eq:contrast2}. 
By Taylor's formula, 
$$ {\int_0^1} \mathcal{C}_n(\theta_0+ {w}(\hat \theta_n - \theta_0)) d{w}\; \;  \mathcal{E}_n = \mathcal{D}_n$$
where
\begin{equation*}
\mathcal{C}_n(\theta) = \left[ \begin{array}{cc}
\frac1{n\Delta_n} \frac{\partial^2}{\partial \varphi^2}\mathcal{L}_n(\theta) & \frac1{n\sqrt{\Delta_n}}  \frac{\partial^2}{\partial \varphi\sigma}\mathcal{L}_n(\theta)\\
 \frac1{n\sqrt{\Delta_n}}  \frac{\partial^2}{\partial \varphi\sigma}\mathcal{L}_n(\theta) & \frac1n  \frac{\partial^2}{\partial \sigma^2}\mathcal{L}_n(\theta)\\
 \end{array}\right], 
 \end{equation*}
\begin{equation*}
\mathcal{E}_n = \left[\begin{array}{c}
 \sqrt{n\Delta_n} (\hat \varphi_n - \varphi_0)\\
 \sqrt{n} (\hat \sigma_n - \sigma_0)\\
 \end{array}
 \right],  \quad 
  \mathcal{D}_n = \left[\begin{array}{c}
 - \frac1{\sqrt{n\Delta_n}} \frac{\partial}{\partial \varphi}\mathcal{L}_n(\theta_0)\\
  - \frac1{\sqrt{n}} \frac{\partial}{\partial \sigma}\mathcal{L}_n(\theta_0)\\
 \end{array}
 \right]
 \end{equation*}
 Lemmas \ref{lemma8}-\ref{lemma9}-\ref{lemma10} and \ref{lemma:Th45Gloter} allow to prove that
\begin{equation}
\mathcal{D}_n \stackrel {\mathcal{D}}{\rightarrow}\mathcal{N}\left(0,\left[\begin{array}{cc}
4 \int \frac{(\partial_\varphi B_2)^2}{ \Gamma^2}(\cdot ; \theta_0) \nu(dx) & 0\\
0 & 2 \int (\frac{\partial_\sigma \Gamma^2}{\Gamma^2})^2(\cdot ; \theta_0) \nu(dx) \\
\end{array}
\right]\right)
\end{equation}
\citep[see][for more details]{Kessler1997}. From Lemmas
\ref{lemma8}-\ref{lemma9} follows 
\begin{equation*}
 \mathcal{C}_n(\theta_0) \rightarrow C:= \left[\begin{array}{cc}
2 \int \frac{(\partial_\varphi B_2)^2}{ \Gamma^2}(\cdot ; \theta_0) \nu(dx) & 0\\
0 &  \int (\frac{\partial_\sigma \Gamma^2}{\Gamma^2})^2(\cdot ; \theta_0) \nu(dx) \\
\end{array}
\right]
\end{equation*}
 Using the consistency of $\hat \theta_n$, we obtain the result. 
  $\hfill \Box$

\subsection{Proof of the asymptotic normality of $(\hat \psi)$}

\paragraph{Proof of Theorem \ref{prop:normality2}.} 
 Denote   $\hat \theta_n=(\hat \psi,  \varphi_0, \sigma_0)$.  Let
 $\mathcal{L}_n(\theta) =  \ell_n(\psi, \sigma)$ from
 \eqref{eq:contrast3}. 
By Taylor's formula, 
$$  {\int_0^1}  \mathcal{C}_n(\theta_0+ {w}(\hat \theta_n - \theta_0)) d{w}\; \;  \mathcal{E}_n = \mathcal{D}_n$$
where
\begin{equation*}
\mathcal{C}_n(\theta_0) = \frac{\Delta_n}{n} \frac{\partial^2}{\partial
  \psi^2}\mathcal{L}_n(\theta) ,  \quad 
\mathcal{E}_n = \sqrt{\frac{n}{\Delta_n}} 
 (\hat \psi_n - \psi_0),  \quad 
  \mathcal{D}_n = - \sqrt{\frac{\Delta_n}{n}} \frac{\partial}{\partial \psi}\mathcal{L}_n(\theta_0).
 \end{equation*}
Lemma \ref{lemma:Th45GloterV} yields
\begin{equation}
\mathcal{D}_n \stackrel {\mathcal{D}}{\rightarrow}\mathcal{N} \left (
  0, 12 \int
\frac{(\partial_\psi B_1)^2}{ \Gamma^2 (\partial_u a)^2}(\cdot ; \theta_0)
\nu(dx) \right )
\end{equation}
and Lemmas \ref{lemma10} and \ref{lemma8} yield
\begin{equation*}
 \mathcal{C}_n(\theta_0) \rightarrow C:= 6 \int \frac{(\partial_\psi
   B_1)^2}{ \Gamma^2 (\partial_u a)^2}(\cdot ; \theta_0) \nu(dx). 
\end{equation*}
 Using the consistency of $\hat \theta_n$, we obtain the result. 
  $\hfill \Box$

\section{Supplementary material: details on SAEM-SMC algorithm}

\subsection{Assumptions for 
  convergence of moment equation}\label{App:assumptionsmoment}

{The assumptions for the moment equation \eqref{eq:generator} to hold
are as follows. For an ergodic diffusion with invariant measure with
Lebesgue density $\mu$, let $\Phi$ be the class of real functions $f$
defined on the state space $\mathcal{X}$ that are twice continuously
differentiable, square integrable with respect to $\mu$, and satisfy that 
\begin{itemize}
\item $\int_{\mathcal{X}} (Lf(x))^2 \mu (x) dx < \infty$
\item $ \sum_{i,j=1}^{p+1}\int_{\mathcal{X}} \partial_{x_i}
  f(x) \partial_{x_j} f(x)  C_{i,j}(x)  \mu (x) dx < \infty$
\end{itemize}
Then \eqref{eq:generator} holds for the diffusion process \eqref{GeneralModel}, if
it is is ergodic, $f$ is $2(k + 1)$ times continuously 
differentiable, and $Lf \in \Phi$ for $i = 0,...,k$. 
}

\subsection{Assumptions for SAEM convergence}\label{App:assumptions}

\begin{itemize}
\item[\textbf{(M2)}] The functions $\psi(\theta)$ and $\nu(\theta)$ are twice continuously differentiable on $\Theta$. 
\item[\textbf{(M3)}] The function $\bar s:\Theta\longrightarrow
  \mathcal{S}$ defined by 
$ \bar s (\theta) = \int S(\vsmall, \usmall) \pDelta(\usmall|\vsmall;\theta) d\vsmall\, d\usmall$
is continuously differentiable on $\Theta$.
\item[\textbf{(M4)}] The function $\ell_\Delta(\theta) = \log \pDelta(\vsmall,\usmall,\theta)$ is continuously differentiable on $\Theta$ and 
$ \partial_\theta \int \pDelta(\vsmall, \usmall;\theta) d\vsmall\, d\usmall = \int \partial_\theta \pDelta(\vsmall, \usmall;\theta) d\vsmall\, d\usmall .$
\item[\textbf{(M5)}] Define $L:\mathcal{S}\times\Theta\rightarrow \mathbb{R}$ by
$ L(s,\theta) = -\psi(\theta)+\langle s,\nu(\theta)\rangle$. 
There exists a function $\hat \theta:\mathcal{S}\rightarrow \Theta$ such that
$\forall \theta \in \Theta, \, \forall s \in \mathcal{S}, \, L(s,\hat\theta(s))\geq L(s,\theta).$
\item[\textbf{(SAEM1)}] The positive decreasing sequence of the
  stochastic approximation  $(a_m)_{m \geq 1}$ is such that $\sum_{m}
  a_m = \infty$ and $\sum_{m}a^2_m < \infty$. 
\item[\textbf{(SAEM2)}] $\ell_{\Delta} :\Theta\rightarrow \mathbb{R}$ and $\hat\theta:\mathcal{S}\rightarrow \Theta$ are $d$ times differentiable, where $d$ is the dimension of $S(\vsmall, \usmall)$.
\item[\textbf{(SAEM3)}] For all $\theta \in \Theta$, $\int ||
  S(\vsmall, \usmall)||^2\, \pDelta(\usmall|\vsmall;\theta)d\usmall< \infty$ and the
  function $\Gamma(\theta)=Cov_\theta(S(\cdot,\Uton))$ is continuous,
  where the covariance is under the conditional distribution $\pDelta
  (\Uton | \Vton; \theta)$.
\item[\textbf{(SAEM4)}] Let $\{\mathcal{F}_m\}$ be the increasing
  family of $\sigma$-algebras generated by the random variables $s_0,
  \Uton^{(1)}$, $\Uton^{(2)}, \ldots, \Uton^{(m)}$.  For any positive
  Borel function $f$,  
$\EDelta(f(\Uton^{(m+1)})|\mathcal{F}_m) = \int f(\usmall)\pDelta(\usmall|\vsmall,\hthetam) d\usmall$. 
	\item[\textbf{(SMC1)}] The number of particles $K$ used at
          each iteration of the SAEM algorithm varies along the
          iteration: there exists a function $g(m) \rightarrow \infty$
          when $m \rightarrow \infty$ such that
          $K(m) \geq g(m) \log(m)$. 
\item[\textbf{(SMC2)}] The function $S$ is bounded uniformly in $u$.
\item[\textbf{(SMC3)}] The functions $\pDelta(\V_i|\U_i, \V_{i-1},
  \U_{i-1}; \theta)$ are bounded uniformly in $\theta$. 

\end{itemize}

 \subsection{Sufficient statistics of the HO model}\label{sec:suff_stat_HO}
 
We detail the sufficient statistics for the HO model. 
Let us denote $Y_i =  V_{i+1}-V_i - U_i$. There are 6 statistics:
\begin{eqnarray*}
S_1 &=&\frac1{\Delta^5} \sum_{i=0}^{n-1} \left(-\frac{\Delta^3}2 U_i Y_u + \frac{\Delta^3}6(U_{i+1}-U_i)V_i + \frac{\Delta^4}3(U_{i+1}-U_i)U_i \right)\\
S_2 &=&\frac1{\Delta^5} \sum_{i=0}^{n-1} \left( -\Delta Y_i^2 + \frac23 \Delta^2 Y_i(U_{i+1}-U_i) + \frac{\Delta^3}6(U_{i+1}-U_i)U_i \right)\\
S_3 &=& \frac2{\Delta^5} \sum_{i=0}^{n-1} \left(\frac{\Delta^4}{12} V_i^2+\frac{\Delta^5}{12}U_iV_i+\frac{\Delta^6}{12}U_i^2 \right)\\
S_4 &=&\frac2{\Delta^5} \sum_{i=0}^{n-1} \left(\frac{\Delta^2}3 Y_i^2+\frac{\Delta^4}{12}U_i^2 +\frac{\Delta^3}6Y_iU_i \right) \\
S_5 &=&\frac1{\Delta^5} \sum_{i=0}^{n-1} \left(\frac{\Delta^3}6Y_iV_i + \frac{\Delta^4}6 U_iV_i + \frac{\Delta^4}3U_iY_i + \frac{\Delta^5}{12}U_i^2 \right) \\
S_6 &=&\frac1{\Delta -\Delta^2 +\Delta^3/3} \sum_{i=0}^{n-1} \left(U_{i+1}-U_i - \Delta (-D V_i - \gamma U_i) \right)^2
\end{eqnarray*}

Then the maximisation step and the updates of the parameters are as follows:
\begin{eqnarray*}
\hat D_m &=& \frac{S_2S_5-S_1S_4}{S_3S_4-S_5^2}\\
\hat \gamma_m &=& \frac{S_1S_5-S_2S_3}{S_3S_4-S_5^2}\\
\hat \sigma^2_m&=& \frac{S_6}{n\Delta}
\end{eqnarray*}

\end{document}